\documentclass[]{aastex631}
\usepackage{ulem}

\def\cR{{\mathcal R}}
\def\q{\qquad}

\def\gx{{GX~Simulator }}
\newcommand{\bfemph}[1]{\textbf{\emph{#1}}}

\begin{document}

\title{Data-Constrained Solar Modeling  with GX Simulator}

\correspondingauthor{Gelu M. Nita}
\email{gelu.m.nita@njit.edu}

\author[0000-0003-2846-2453]{Gelu M. Nita}
\affiliation{New Jersey Institute of Technology, Newark 07102-1982, NJ, USA}

\author[0000-0001-5557-2100]{Gregory D. Fleishman}
\affiliation{New Jersey Institute of Technology, Newark 07102-1982, NJ, USA}

\author[0000-0001-8644-8372]{Alexey A. Kuznetsov}
\affiliation{Institute of Solar-Terrestrial Physics, Irkutsk 664033, Russia}

\author[0000-0002-1107-7420]{Sergey A. Anfinogentov}
\affiliation{Institute of Solar-Terrestrial Physics, Irkutsk 664033, Russia}

\author[0000-0002-5453-2307]{Alexey G. Stupishin}
\affiliation{Saint Petersburg State University, 7/9 Universitetskaya nab., St. Petersburg, 199034 Russia}

\author[0000-0002-8078-0902]{ Eduard P. Kontar}
\affiliation{University of Glasgow, Glasgow, G12 8QQ, UK}

\author[0000-0002-5476-2794]{Samuel J. Schonfeld}
\affiliation{Institute for Scientific Research, Boston College, Newton, MA 02459, USA}

\author[0000-0003-2255-0305]{James A. Klimchuk}
\affiliation{NASA Goddard Space Flight Center, Greenbelt, MD 20771, USA}

\author[0000-0003-2520-8396]{Dale E. Gary}
\affiliation{New Jersey Institute of Technology, Newark 07102-1982, NJ, USA}



\begin{abstract}

To facilitate the study of solar active regions and flaring loops, we have created a modeling framework, the freely distributed \gx IDL package, that combines 3D magnetic and plasma structures with thermal and non-thermal models of the chromosphere, transition region, and corona. The package has integrated tools to visualize the model data cubes, compute multi-wavelength emission maps from them, and quantitatively compare the resulting maps with observations. Its object-based modular architecture, which runs on Windows, Mac, and Unix/Linux platforms, offers capabilities that include the ability to either import 3D density and temperature distribution models, or to assign numerically defined coronal or chromospheric temperatures and densities, or their distributions to each individual voxel. \gx can apply parametric heating models involving average properties of the magnetic field lines crossing a given voxel, as well as compute and investigate the spatial and spectral properties of radio, (sub-)millimeter, EUV, and X-ray emissions calculated from the model. The application integrates FORTRAN and C++ libraries for fast calculation of radio emission (free-free, gyroresonance, and gyrosynchrotron emission) along with soft and hard X-ray and EUV codes developed in IDL. To facilitate the creation of models, we have developed a fully automatic model production pipeline that, based on minimal users input, downloads the required SDO/HMI vector magnetic field data and (optionally) the contextual SDO/AIA images, performs potential  or nonlinear force free field extrapolations, populates the magnetic field skeleton with parameterized heated plasma coronal models that assume either steady-state or impulsive plasma heating, and generates non-LTE density and temperature distribution models of the chromosphere that are constrained by photospheric measurements. The standardized models produced by this pipeline may be further customized through a set of interactive tools provided by the graphical user  interface. Here we describe the GX Simulator framework and its applications.

\end{abstract}


\section{Introduction}
\label{section:intro}
The fundamental problems of modern solar physics require analysis of multiple vast data sets obtained with a multitude of ground- and space-based instruments. The sheer level of complexity in newly available datasets calls for adequate theoretical modeling in order to derive the target physical parameters of a given measurement. Examples include extrapolating the photospheric magnetic field data from optical observations, or deducing the distribution of thermal coronal plasma from extreme ultraviolet (EUV) observations. Larger caliber theoretical and modeling efforts are needed to meaningfully combine and cross-validate multiple data sets. These data come from space missions, e.g Helioseismic and Magnetic Imager \citep[HMI;][]{Scherrer2012} onboard the Solar Dynamic Observatory \citep[SDO;][]{sdo},  and ground-based high-resolution optical and infrared (IR) instruments such as Goode Solar Telescope \citep[GST;][]{gst,nst} and Daniel K. Inouye Solar Telescope \citep[DKIST;][]{atst,dkist}. Fundamental enhancements of theory and modeling are demanded to fully exploit new observational windows such as microwave and millimeter-wave imaging spectropolarimetry data from the Expanded Owens Valley Solar Array \citep[EOVSA;][]{Nita2016,Gary2018}, the Siberian Radio Heliograph \citep[SRH;][]{srh1, srh2} and the Atacama Large Millimeter/submillimeter Array (ALMA), in addition to more traditional X-ray and EUV data, e.g the Reuven Ramaty High-Energy Solar Spectroscopic Imager \citep[RHESSI;][]{rhessi}, the Atmospheric Imaging Assembly\citep[AIA;][]{Lemen2012} onboard SDO, or the Spectrometer/Telescope for Imaging X-rays \citep[STIX;][]{stix}. Dynamic solar phenomena that constitute solar activity either occur in or are sensitive to the physical conditions in the solar corona. The dominant form of energy in the solar corona is magnetic energy; thus, the knowledge of the coronal magnetic field is central for understanding coronal physics. However, there is no observational technique that provides the 3D magnetic vector field over a significant coronal volume. This is why data-constrained modeling of the coronal magnetic field is extremely important.

The \gx modeling framework that we present here is based on a magnetic model (magnetic skeleton), which, once created, can be populated by thermal plasma and nonthermal particles, and then various emissions can be computed from the volume and compared with observations. When all synthesized observables match all available data, the model is proved to be valid.

The challenges that our modeling framework solves are: (i) automated creation of the magnetic model; (ii) addition of an objectively defined thermal structure of the corona and chromosphere; (iii) rigorous calculation of radio, EUV, and X-ray continuum emission from the model; and (iv) provision for model-to-data comparison. To facilitate the creation and manipulation of the models, the tool offers numerous options. Earlier versions of the tool were described by \citet{Nita_etal_2015} for the flare science and \citet{Nita_etal_2018} for the active region (AR) science. This paper summarizes the functionality of those initial versions and describes numerous updates and enhancements of the tool.

\section{The GX Simulator Automatic Model Production Pipeline}
\label{section:pipeline}
\subsection{General Description of the Pipeline Functionality}
To facilitate the use of the \gx modeling package \citep{Nita_etal_2015,Nita_etal_2018}, we have developed a fully automatic model production pipeline (AMPP) that, based on minimal user's input, downloads the required vector magnetic field data produced by the Helioseismic and Magnetic Imager (HMI) onboard the Solar Dynamics Observatory \citep[SDO,][]{Scherrer2012} and (optionally) the contextual Atmospheric Imaging Assembly maps \citep[AIA,][]{Lemen2012}, performs potential  and/or nonlinear force free field (NLFFF) extrapolations, populates the magnetic field skeleton with parameterized heated plasma coronal models that assume either steady-state or impulsive plasma heating, and generates non-LTE density and temperature distribution models of the chromosphere that are constrained by photospheric measurements. The standardized models produced by this pipeline may be further customized through a set of interactive tools provided by the \gx graphical user interface (GUI).

The AMPP sub-module is exposed to the users through a single top-level IDL routine, namely \textbf{gx\_fov2box.pro}, which provides a series of options that may be used to customize its functionality, as detailed in Appendix \ref{appendix:a}, where we provide an AMPP script  to generate a magneto-thermal model for an instance of AR11520 observed on 12-Jul-2012 04:58:26 UT, which we use as an illustrative example in the subsequent sections.

\begin{figure}[!htb]
\centering
\includegraphics[width=0.8\textwidth]{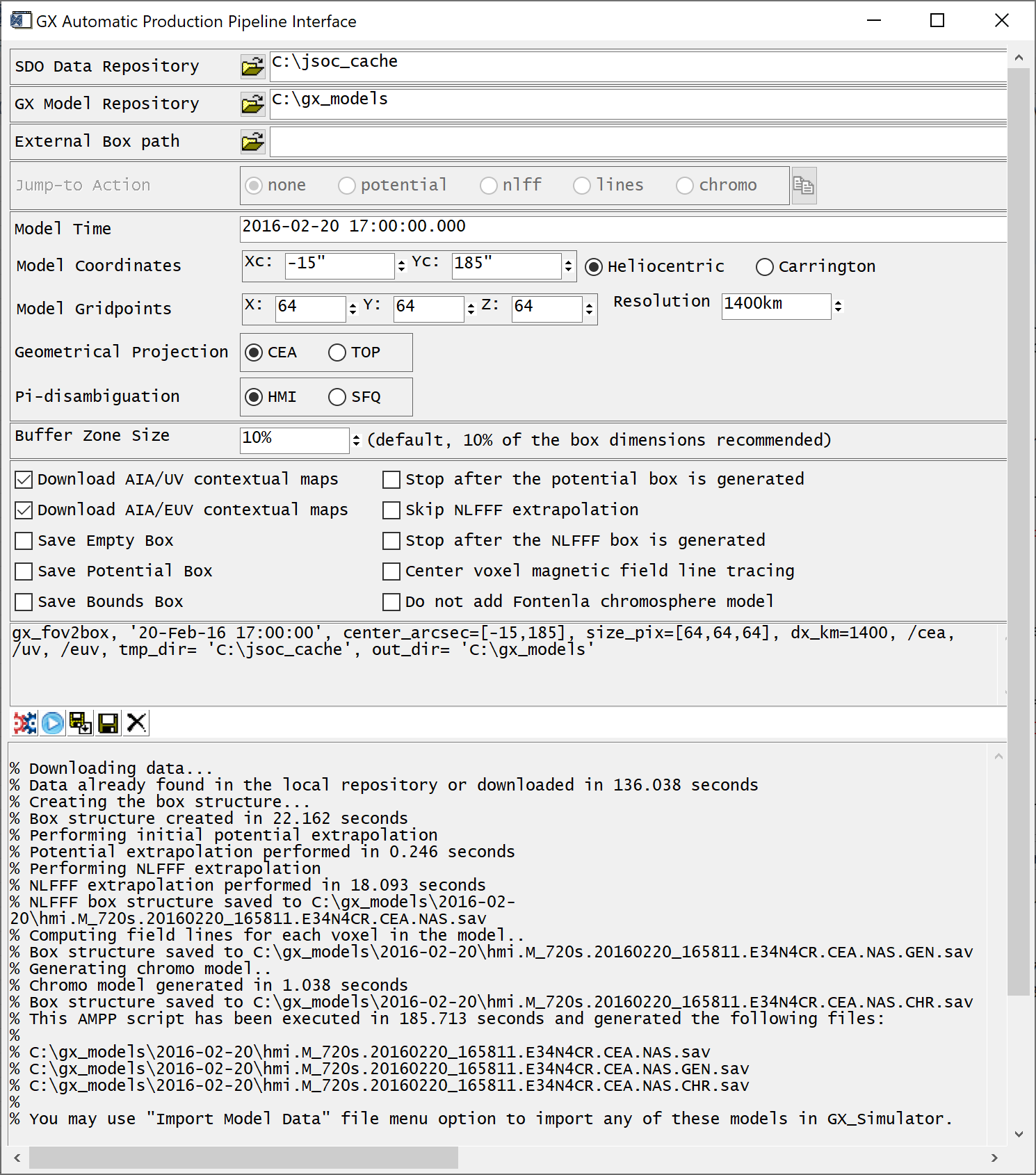}
\caption{
\label{gx_ampp} Snapshot of the gx\_ampp GUI application displaying a set of default settings and corresponding run-time execution messages that match the demo script \textbf{create\_box\_20160220.pro} included in the \textbf{/gx\_simulator/demo/} sub-folder of the \gx distribution.
}
\end{figure}

The \gx package also provides a standalone GUI application, \textbf{gx\_ampp.pro}, which may be used to conveniently generate and run AMPP scripts, as illustrated in Figure \ref{gx_ampp}, which displays a set of default settings and corresponding run-time messages that match the demo script \textbf{create\_box\_20160220.pro} included in the \textbf{/gx\_simulator/demo/} sub-folder of the \gx distribution.

Any interactive change of the input fields of the \textbf{gx\_ampp} GUI  updates the functional \textbf{gx\_fov2box} script, which may be launched from the interface, or copied and run directly from the IDL command line. The \textbf{gx\_ampp} GUI also provides the option of uploading an already existing \gx compatible box structure (such as a potential or NLFFF extrapolation box), which may be used as a starting point for adding  properties to the model, such as optional magnetic field tracing parameters and/or chromosphere models, as described in the subsequent sections.

\begin{figure}[!htb]
\includegraphics[width=1\columnwidth]{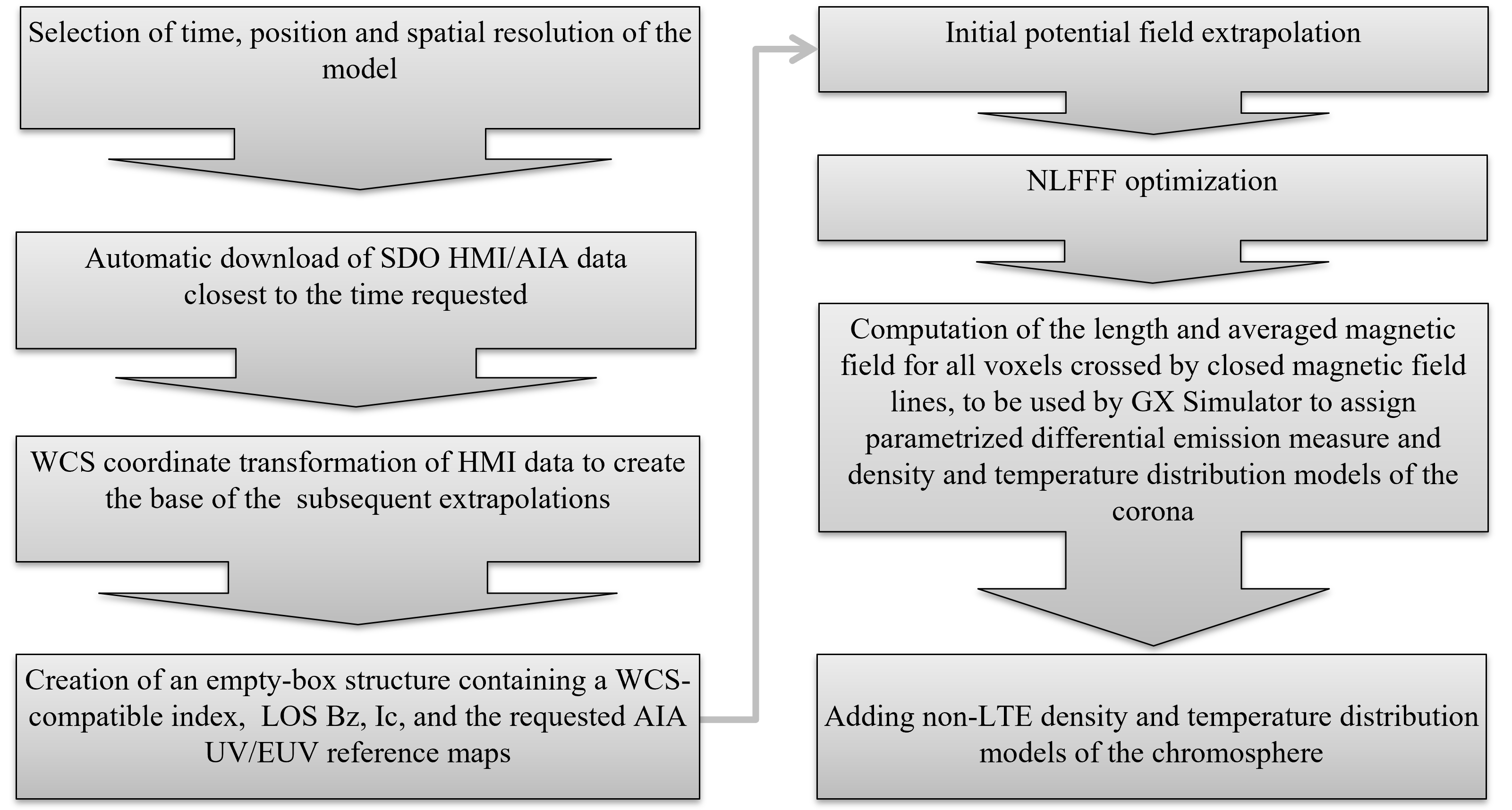}
\caption{
\label{pipeline} \gx Automatic Model Production Pipeline workflow.
}
\end{figure}

\begin{figure}[!htb]
\centering

\includegraphics[width=0.45\columnwidth]{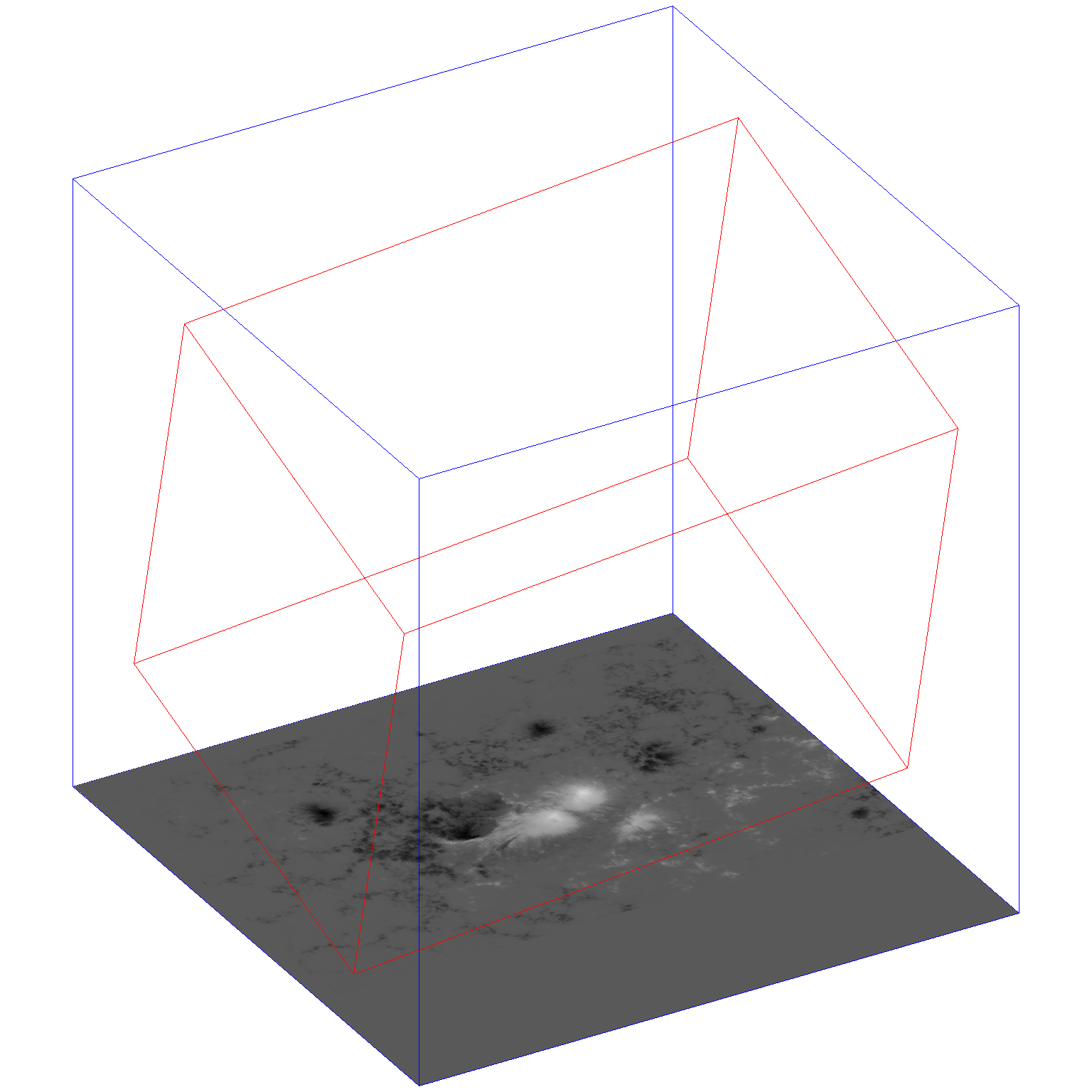}(a)
\includegraphics[width=0.45\columnwidth]{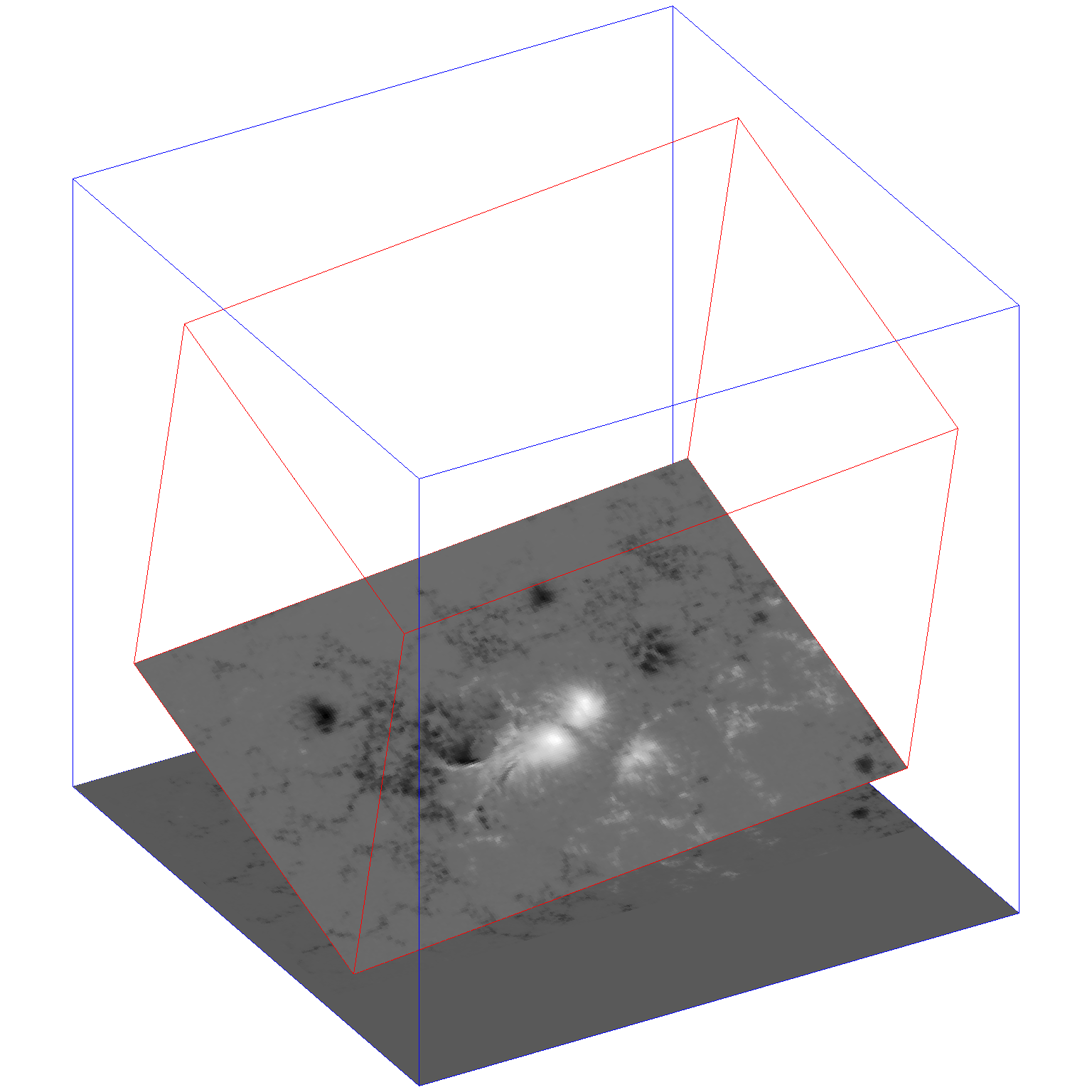}(b)
\includegraphics[width=0.45\columnwidth]{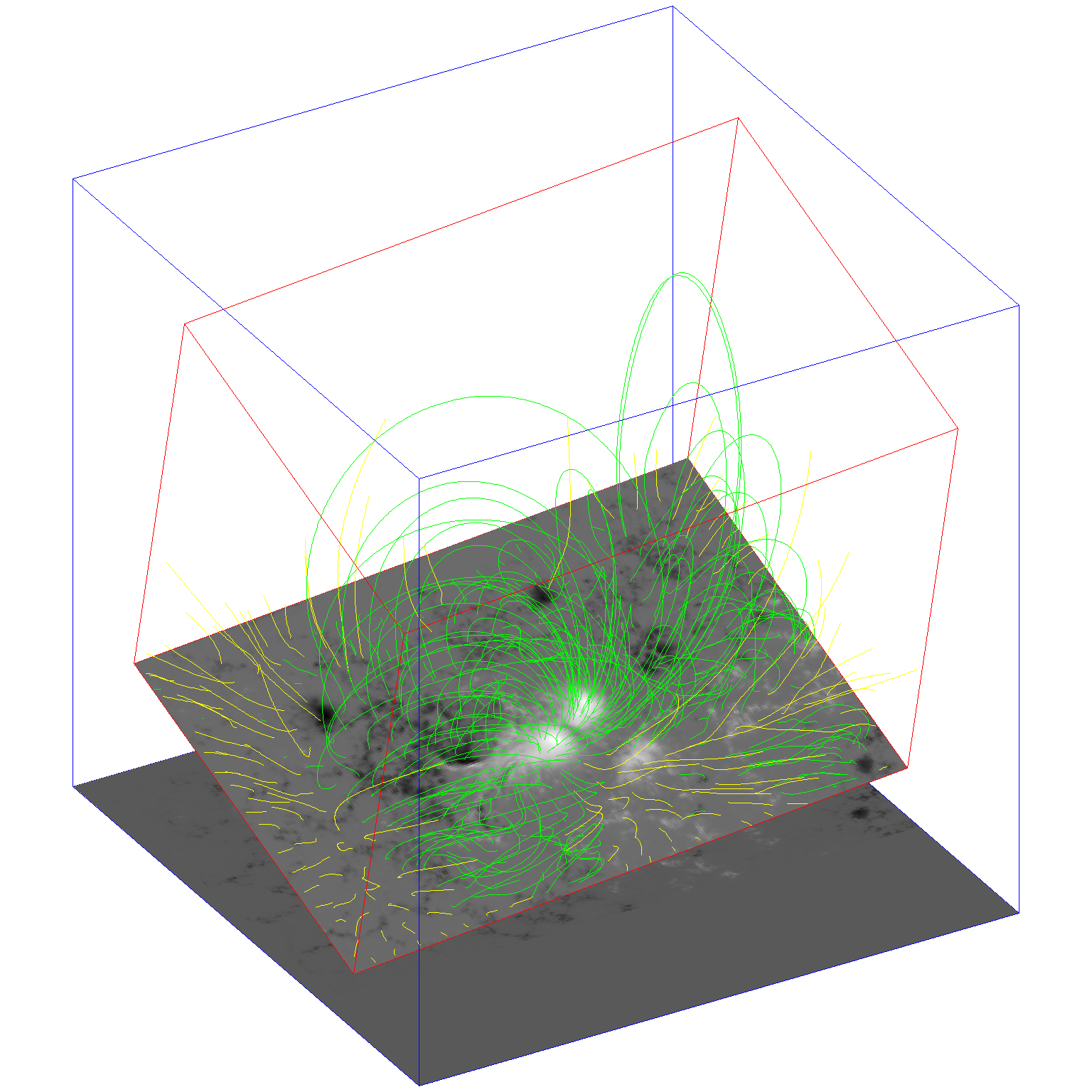}(c)
\includegraphics[width=0.45\columnwidth]{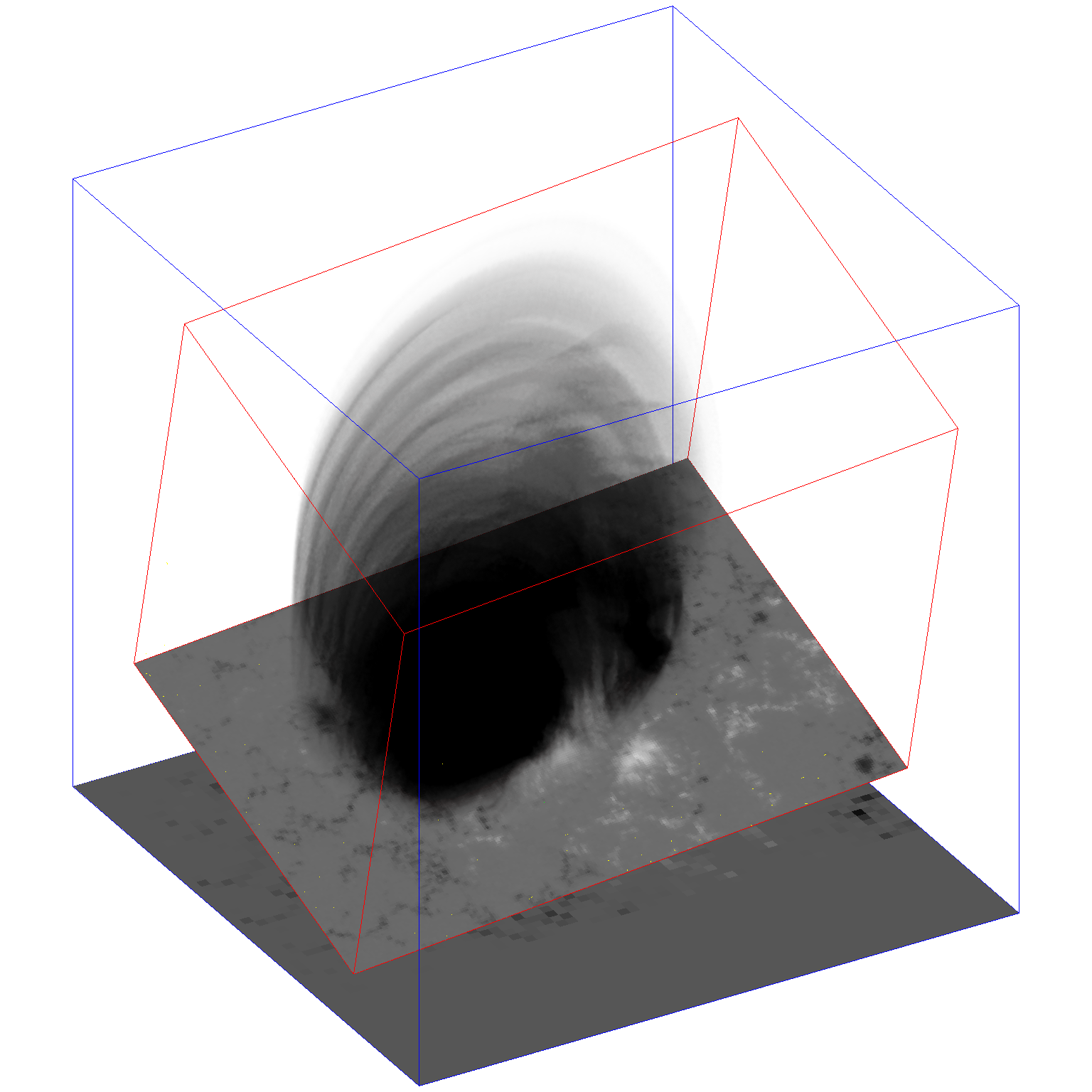}(d)
\caption{
\label{model_ampp}
Snapshots of the AR11520 model generated by the script presented in Appendix \ref{appendix:a} illustrating different stages of the AMPP process. (a) The photospheric LOS magnetic field map, located at the bottom of a rectangular box co-aligned with the observer's LOS (blue lines). The inscribed rectangular box (red lines), which is aligned with the direction normal to the solar surface, defines the 3D volume in which the magnetic field extrapolation is performed. (b) The elements in (a) plus $B_z$, obtained by projecting the photospheric vector magnetic field HMI map onto the bottom boundary.  Two projection options are provided by AMPP: the default cylindrical area projection (CEA), or a simple parallel projection (TOP), selected by using the "Geometrical Projection" radio button shown in Figure~\ref{gx_ampp}. (c) A sub-set of model field lines that do, or do not, close within the 3D model boundaries (green and yellow lines, respectively), illustrating the magnetic connectivity in the model. (d) The coronal temperature distribution along the closed field lines, which corresponds to the parameterized magneto-thermal model defined by the AMPP script (refer to \S\ref{section:corona} for a detailed description). A user-specified hydrostatic model is used for the volume outside these closed field lines.
}
\end{figure}
Figure \ref{pipeline} illustrates the main building blocks and the workflow of the \gx AMPP module and Figure \ref{model_ampp} displays a series of snapshots of the 3D magnetic model produced by the AMPP script provided in Appendix \ref{appendix:a}.
The initialization of an AMPP run, illustrated by the first two blocks of the workflow diagram shown in Figure~\ref{pipeline}, requires only the time, field of view (FOV), height, and desired spatial resolution of the model. The time and location input parameters are used by the AMPP to identify and download the available SDO HMI/AIA maps closest to the requested time, after checking the specified local repository in case they were already downloaded during a previous AMPP run.

These input SDO HMI/AIA data products are used to prepare a data structure and boundary conditions needed to perform the subsequent AMPP tasks (blocks 3 and 4 of Figure~\ref{pipeline}). To do so, the AMPP creates an initial empty volume structure on top of the photospheric vector magnetogram boundary conditions that are prepared by performing Carrington-Heliographic or Helioprojective-Cartesian projection \citep{Thompson2006}, as illustrated in panels (a) and (b) of Figure \ref{model_ampp}. Unlike the standard, general-purpose Spaceweather HMI Active Region Patch \citep[SHARP,][]{Bobra2014} data products routinely used as boundary conditions by other magnetic field reconstruction packages, the AMPP boundary condition maps are exactly centered on the user-requested FOV, which minimizes to the maximum extent possible the unavoidable projection effects.

In the next stage (blocks 5 and 6 of Figure~\ref{pipeline}), the AMPP applies the method described in \S\ref{POT} to produce an initial potential field extrapolation 3D structure, which is used in the next stage as an initial condition for generating a NLFFF model using the optimization code described in \S\ref{NLFFF}.
If not explicitly disabled by the user, the next AMPP block computes the averaged magnetic field $\langle B\rangle$ and length $L$ of the potential or NLFFF magnetic field lines crossing each volume voxel. This enables \gx to interactively dress the magnetic skeleton with a parameterized thermal structure, as detailed in \S\ref{section:corona}. Panels (c) and (d) in Figure \ref{model_ampp} illustrate a series of magnetic field lines and their associated magneto-thermal structure corresponding to the NLFFF magneto-thermal model generated by the AMPP script presented in Appendix \ref{appendix:a}. Finally, if not explicitly disabled by the user, the last block of the AMPP workflow diagram replaces the bottom layers of the potential or NLFFF model with a non-LTE chromosphere model, as described in \S\ref{section:chromo}.

As illustrated in Figure \ref{gx_ampp},  there is a set of optional keyword switches to skip some optional execution blocks and/or to save, in addition to the final model, any intermediary models generated by the workflow. Thus, to help distinguish these AMPP products without the need to inspect the output files, we have adopted a file naming convention that combines a series of distinctive tags that uniquely identify each type of model, as listed in Table \ref{tab:file_types}. For example, an AMPP output file tagged as \textbf{``.NAS.GEN.CHR."} would indicate a NLFFF model augmented by adding $\langle B\rangle$-$L$ properties and a non-LTE chromosphere, while \textbf{``.POT.GEN.CHR."} would denote that the same additional properties were added to a potential magnetic field model, if the user chooses to skip the NLFFF optimization block. Any IDL structure produced by the AMPP contains a string tag named ``EXECUTE," which provides an exact copy of the execution script used to generate it.

\begin{deluxetable}{p{0.08\linewidth}|p{0.8\linewidth}}[h!]
\tablecaption{\gx filename extension naming convention  \label{tab:file_types}}
\tablehead{\colhead{Tag}&\colhead{Model Type} }
\startdata
.NONE. &
   An empty-box IDL structure that contains all geometrical information and context SDO/AIA maps requested, as well as properly sized zeroed arrays ready to store the Cartesian components of the magnetic field model not yet generated, as described in \S\ref{box}\\
.POT. &
An IDL structure containing a true potential solution based on only the $B_z$ base map component, as described in \S\ref{POT}.\\
.BND. &
An IDL structure containing potential solution except for the bottom layer, which is replaced by the observed $B_y$ and $B_z$ components – to be used as initial conditions for the following NLFFF optimization step.\\
.NAS. &
An IDL box structure filled with a non linear force free field magnetic field model \citep{Stupishin2020}, as described in \S\ref{NLFFF}\\
.GEN. &
An IDL box structure containing the length $L$ and averaged magnetic field $\langle B\rangle$ along the field lines crossing a given volume voxel. These additional parameters are ready to be used for the purpose of adding a parameterized heated plasma coronal models that assume either steady-state or impulsive plasma heating, as described in \S\ref{section:corona}\\
.CHR. &
An IDL box structure having the bottom layers of uniform-height replaced by a non-uniform height, non-LTE density and temperature distribution model of the chromosphere that is constrained by photospheric measurements (Fontenla et al. 2009), as described in \S\ref{section:chromo}. \\
\enddata
\end{deluxetable}

\subsection{AMPP NLFFF Magnetic Field Models}
\subsubsection{Preparation of the boundary and initial conditions for the NLFFF extrapolation\label{box}}
\begin{figure}[!htb]
\centering
\includegraphics{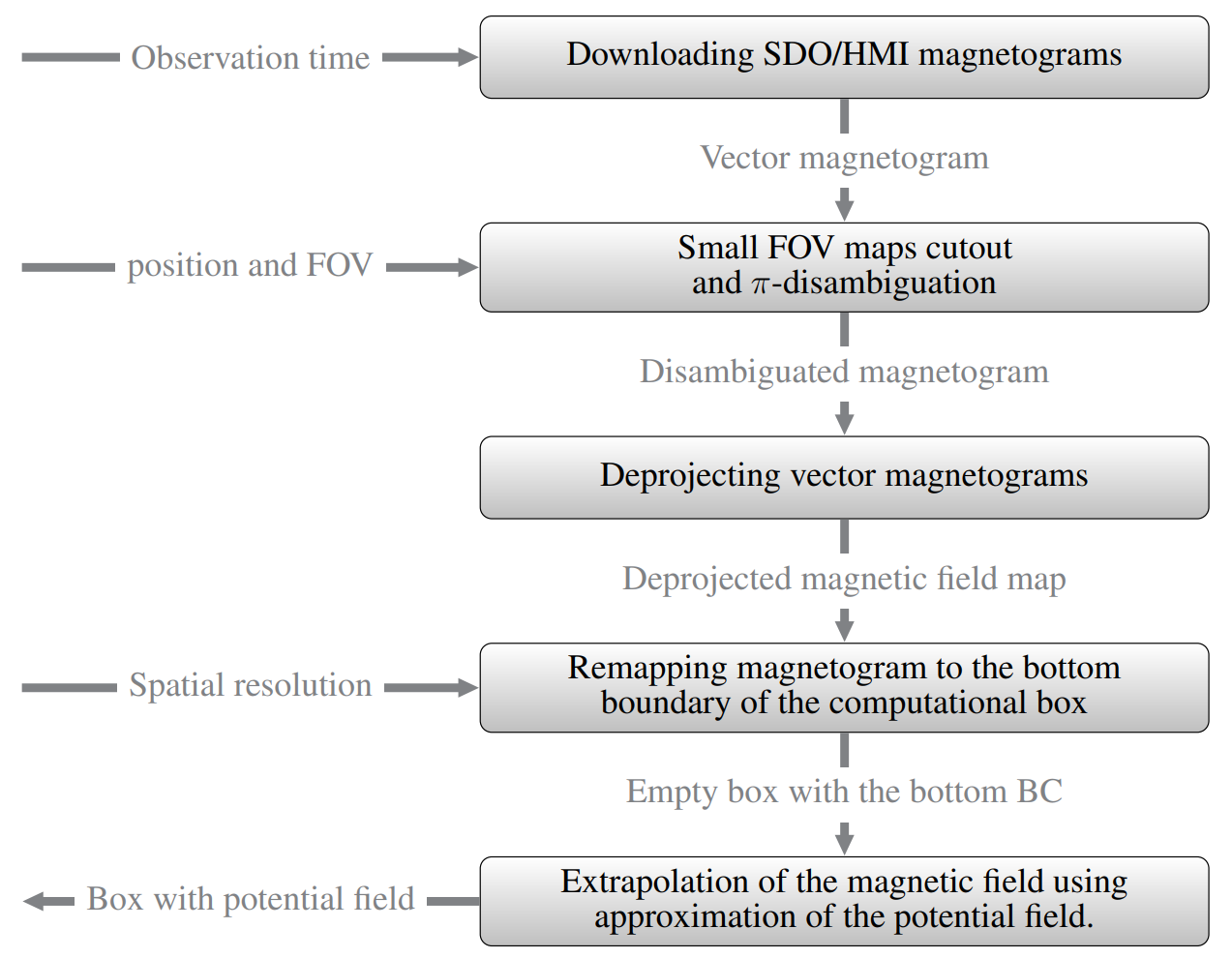}



    \caption{Production of initial and boundary conditions flowchart. Gray boxes represent individual steps of the pipeline, while intermediate data products are shown as arrows with labels.
    }
    \label{fig:AMPP_initial_conditions}
\end{figure}

The first stage of the AMPP is the production of initial and boundary conditions for the subsequent NLFFF extrapolation.
The user provides  AR coordinates, observation time,  and the size and spatial resolution for the resulting 3D data cube.
Then the pipeline will download required data and produce a data cube with the photospheric measurements of the magnetic field vector in its bottom layer. The rest of the volume will be filled with the extrapolated potential field. These operations are fully automated and do not require any additional actions from the user.
The flowchart of production of the initial and boundary condition is shown in Figure \ref{fig:AMPP_initial_conditions}.
The individual steps of this AMPP stage are described below.

Firstly, the pipeline script automatically downloads, from the JSOC data processing center, SDO/HMI vector magnetograms (data series:hmi.B\_720s) taken at the time closest to the time requested by a user.
For further processing, the limited field of view (FOV) maps are cut out from the full Sun magnetograms. The precomputed $\pi$-disambiguation provided by the JSOC data processing center is applied using the \texttt{HMI\_DISAMBIG} routine from the Solar Soft library.
In the case of disambiguation artifacts, there is an option to perform $\pi$-disambiguation with the Super Fast and Quality azimuth disambiguation library \citep[SFQ, ][]{sfq}, also
known as the new disambiguation method (NDA), which is supplied as a part of the AMPP. Although both methods,which may be interchanged by using the HMI/SFQ switch, work comparably well \citep{Fleishman2017b}, in some cases, especially for near limb observations, the SFQ library may provide better results than the standard HMI disambiguation \citep{sfq}.

After the disambiguation, the vector magnetic field map is deprojected from the LOS coordinate system to the spherical components  $B_\phi$, $B_\theta$, and $B_r$ using the \texttt{ HMI\_B2PTR} procedure from the SDO/HMI package in solar soft. These components will become $B_x$, $-B_y$, and $B_z$ components in the cartesian coordinate system of the computational box.

At the next step, the deprojected magnetic field components are remapped to the local coordinate system of a computational box.
Since the current version of AMPP uses cartesian coordinates, the magnetic field maps are projected from the spherical surface of the Sun to the flat bottom of the box.
The AMPP supports two projections: top view which is a simple parallel projection and  cylindrical equal area (CEA) projection.
The latter is the default option and preferable for extrapolation purposes since it preserves the area of magnetic elements and, hence, the magnetic flux is not changed by the projection effects.
While performing coordinate transformation AMPP relies on a WCS general purpose library that is supplied by the Solar SoftWare (SSW) repository.
To improve the quality of remapping, we use cubic interpolation when the requested resolution of the computational box is higher or comparable with the pixel size of the available magnetograms.
In the opposite case, when the spatial resolution of the computational box is lower than the resolution of a magnetogram, we use an over-sampling antialising technique by dividing every computational pixel into 8 subpixels.
The resulting  magnetic field components are then converted to the computational resolution by direct summation of the values interpolated to subpixels.

After remapping, the map of a deprojected magnetic field is placed in the computational box as a bottom layer, the rest of the computational box consisting of zeroed arrays, ready to store the Cartesian components of the magnetic field model not yet generated. This geometrical structure, which may be optionally saved to disk as a file with \textbf{``.NONE."} tag, is forwarded to the next stage of the AMPP.

\vspace{12pt}
\subsubsection{Potential Field Initialization of the AMPP Model\label{POT}}

During this stage of the AMPP process, the empty box volume is filled with a potential field solution obtained from the normal component of the magnetic field at the lower boundary using the Fast Fourier Transform (FFT) method described in \citet{1981A&A...100..197A}.
The FFT solution for the potential field problem implies periodic, flux-balanced boundary conditions at lateral boundaries, which is not realistic. To simulate a more appropriate ``open" boundaries, we expand the computational domain ($L_{x,y}$) by $L_{x,y}/2$ in each direction. Then, the normal component of the field at the lower boundary is padded with a constant, generally non-zero value. This value is computed such as the total signed magnetic flux from the added areas perfectly compensates the unbalanced flux at the original lower boundary. The final potential field solution is then obtained by cutting out from the expanded domain and is then used as initial condition for the NLFFF extrapolation.

\vspace{12pt}
\subsubsection{NLFFF Reconstruction and Magnetic Field Line Tracing Dynamic Link Library}

\label{NLFFF}
The NLFFF reconstruction code employed by AMPP was developed in C++ using multithreaded functionality. Its source, the compilation scripts designed for Windows~and Linux platforms, a set of compiled libraries for both platforms, and their calling IDL wrappers are included in the \gx SSW distribution package, and are automatically updated from their independently maintained GitHub development repository\footnote{  \href{https://github.com/Alexey-Stupishin/Magnetic-Field_Library}{Magnetic-Field\_Library}}. In addition, the package may be directly downloaded from a Zenodo\copyright~digital repository
\citep{Stupishin2020}\footnote{\href{https://doi.org/10.5281/zenodo.3896222}{Magnetic Field Library: NLFFF and magnetic lines}}.
Since the general platform compatibility of the pre-compiled Linux library is not guaranteed, the AMPP  automatically invokes the distributed source code to compile and save a local copy\footnote{The user may invoke the IDL command line {``print, gx\_libpath('nlfff')"} to retrieve the location of the  shared library , or  {``print, gx\_libpath('nlffff',/update)"}, to also request a new compilation of the library, provided that a g++ compiler is installed on the system.} of the shared library on its first call on a Linux platform, which is used in all subsequent calls.

This NLFFF reconstruction code follows the development proposed by \citet{2000ApJ...540.1150W} and \citet{2004SoPh..219...87W}. The basic idea is to reduce the Lorentz force in the coronal volume (i.e., to eliminate transverse electric currents) and reduce the field divergence as much as possible by minimization of the functional

\begin{equation}\label{Eq_nlfff_func}
L=\int\limits_{V} \left[B^{-2}\left[ [\nabla\times\mathbf{B}]\times\mathbf{B}\right]^2 + |\nabla\cdot\mathbf{B}|^2\right] w(x,y,z) dV,
\end{equation}
where the first term, ($B^{-2}\left[ [\nabla\times\mathbf{B}]\times\mathbf{B}\right]^2$), represents the Lorentz force, the second one, ($|\nabla\cdot\mathbf{B}|^2$), evaluates the field divergence, while $w(x,y,z)$ is a ``weight'' function. The weight function is intended to diminish the influence of uncertainties of the field at the side and top boundaries, and can be adjusted by the user. By default, the weights are constant (=1) in the entire volume except for $10\,\%$ of the length of each dimension on each side, where the weights decrease to zero on the boundaries following a cosine function (the bottom boundary is not weighted, because it is set to the observed photospheric field).

The initial state of the magnetic field may be inferred from several reasonable approaches. By default, the algorithm uses the preliminary potential field reconstruction described in \S\ref{POT}. Alternatively, the NLFFF reconstruction may be started from an AMPP-compatible geometrical box pre-filled with an initial magnetic field configuration obtained by any means, from which the boundary conditions are also inferred with or without buffer zones, as indicated by the user.

If one considers the magnetic field as a function of the conditional evolution parameter $t$, $\mathbf{B}(x,y,z,t)$,  the evolution of the functional may be estimated by computing $\partial \mathbf{B}/\partial t$ at $i^{th}$ step ($L_i$) and modifying $\mathbf{B}$ for the next $(i+1)^{th}$ step as $\mathbf{B}_{i+1} = \mathbf{B}_{i} + (\partial \mathbf{B}/\partial t) \Delta t$ (where $\Delta t$ is a small evolution step), to get the next functional value $L_{i+1}$. The step size is varied such as to increase the iteration speed, being chosen automatically depending on the speed of convergence: it is increased by 10\,\% at a successful step and decreased by 10\,\% at an unsuccessful one.  

Due to the numerical errors affecting the computation of the functional, the value of the functional may slightly increase at the next iteration even for a small step $\Delta t$, which is allowed by the algorithm up to a $10^{-4}$ relative factor. If the functional increases above this limit, the step is reduced by 10\,\%. The iterations stop when the step becomes too small, i.e. less than 1\,\% of the initial value. In addition, the iterations are terminated if (i) the relative variation of the function for the last 10 iterations is small (less than $5 \cdot 10^{-4}$), or (ii) if the maximum value of $|L_i/L_{i+1} - 1|$ does not exceed $10^{-4}$ during the previous 100 iterations. In such cases, no further significant decrease of the functional is expected, and it is assumed that the current solution is reasonably close to the optimal one.

Another approach used to decrease the computation time is the technique of  ``multigrids'' \citep{2008SoPh..247..269M}. Instead of performing the computations using directly the desired volumetric grid resolution (e.g. 257x257x129), the initial potential field is firstly computed over a smaller resolution grid, let us say, 65x65x33, a first stage NLFFF functional minimization is performed, and then the procedure is repeated twice, increasing the grid resolution at each step (to 129x129x65 and, finally, to 257x257x129), while interpolating the  solution obtained at each step to the next grid resolution and using it as the initial condition for the next step.

The same code may also be used to compute the magnetic field lines passing through each voxel of the volume, or through a set of predefined ``seed-voxels'' (box coordinates).
The lines are computed using the Runge-Kutta-Feldberg algorithm of $4^{th}(-5^{th})$ orders (the code was ported from original FORTRAN implementation, see \citet{Forsythe_1977_CMMC}, and adapted to C++ using multi-thread functionality).
When computation of a set of seeded lines is requested by the user, the code returns each line as a collection of fractional box indices indicating at least one intersection point for each volume element that is intersected. Such lines  may be used for the purpose of visualizing the magnetic field connectivity, as well as for constructing flux-tubes that may be used in flare studies, as described in \S\ref{section:loops}.
\vspace{0.51cm}

\subsection{AMPP Default Coronal Models}
\label{S_corona_ampp}

The computation of the magnetic field lines passing through each voxel of the volume is performed for dressing the magnetic field structure  with a thermal plasma model \citep[][see \S\ref{section:corona} for more details]{Nita2018}. In this case, the code returns the following parameters associated with each voxel of the volume:
\begin{itemize}
\item length of the field line intersecting the voxel,
\item average magnetic field along the line,
\item connectivity with the two boundary voxels associated with the line
\item a flag indicating whether the voxel is intersected by a closed field line (both footpoints at the chromospheric layer are located inside the box) or by an open line (only one footpoint is located inside the box).
\end{itemize}

For a quantitative assessment of the computational speed, the reader may refer to the console messages generated when running the AMPP script presented in Appendix \ref{appendix:a},which was used to generate a $240\times168\times200$ magnetic field cube for an instance of AR11520 on a Windows~10 system equipped with an Intel Xeon E-2286M CPU 2.4 GHz, 8 cores, 64 GB RAM. In this particular case, the NLFFF reconstruction was performed in $\sim250$ seconds and the computation of the lines intersecting all volume voxels was performed in $\sim105$ seconds. For a given size of the computational box the computational time of an NLFFF reconstruction may vary as much as one order of magnitude depending on the complexity of the magnetic field configuration (e.g. isolated sunspot versus a complex AR), while the speed of the full-volume line computation scales roughy linearly with the number of volume elements.
More detailed benchmark tests performed for the purpose of assessing the code accuracy when compared with a ground truth magnetic field model may be found in \citet{Fleishman2017b}, where the code is referred to as the \textbf{AS} NLFFF reconstruction code.\\

\subsection{AMPP Default Chromosphere Models}
\label{section:chromo}
The general approach employed by the AMPP to populate the chromospheric volume, described in detail in \citet{Nita2018}, uses observationally established thresholds to distinguish 7 quiet-Sun (QS) and AR features based on photospheric data, and selects one of a set of 7 corresponding 1D solar atmospheric models proposed by  \citet{Fontenla2009} to fill the chromospheric volume above a particular chromospheric pixel. The 7 feature types comprise 3 QS components, namely internetwork (IN), network lane (NW), and enhanced network (ENW), and 4 AR features: sunspot umbra (UBR), penumbra (PEN), plage (PL), and facula (FA).

The AMPP applies the above selection thresholds automatically, and thus produces the corresponding chromospheric model based on a given HMI limb-darkening-removed white-light map and LOS magnetogram pair, which are downloaded from the SDO/HMI data repository. Thus, using the model mask computed by these means, the pipeline generates a chromospheric volume consisting of a collection of variable height (number of grid-steps) vertical columns, each corresponding to one of the seven chromospheric models associated with the photospheric mask pixel onto which such a column is projected. The chromospheric volume thus generated is then used to replace the bottom layers of the uniformly spaced magnetic skeleton with a composite slab having the minimum thickness needed to contain the variable height chromosphere, and any height-dependent properties of the original volume are interpolated and transferred to the non-uniform chromospheric voxels. The \gx model structures that include such chromosphere models are by default stored on the disk with a filename that includes the \textbf{``.CHR."} tag (although \gx does not rely on this naming convention to recognize the type of models produced by the pipeline).

However, one may choose to skip this step of the model production pipeline, and assign instead a chromosphere represented by a uniform slab  of adjustable height, and constant temperature $T_{chr}$ and density $n_{chr}$, interactively chosen through the \gx GUI, this option being available for any of the \textbf{POT}, \textbf{NAS}, \textbf{NAS.GEN}, or \textbf{POT.GEN} models produced by the pipeline. This simpler option is often appropriate for modeling of flaring loops.\\

\subsection{AMPP User-Adjustable Coronal Models}
\label{section:corona}
The default background corona is populated with an analytically defined  horizontally uniform hydrostatic equilibrium model  \citep{Nita2018}. Alternatively, a field-aligned hydrodynamic model is available to replace the background thermal plasma in voxels on closed loops. This model assumes a heating along the individual magnetic flux tubes defined by the extrapolated field line structure,  The hydrodynamic simulation code called the Enthalpy-Based Thermal Evolution of Loops (EBTEL) {\citep{Klimchuk2008, Cargill2012, Cargill2012a, Barnes2016a, Bradshaw2016, UgarteUrra2017}}, which assumes an impulsive heating (including  a nonoflare mechanism), is used in this case. EBTEL includes the important link between the corona and lower atmosphere in order to realistically model the plasma response to coronal heating.

As illustrated in Figure \ref{pipeline} (see Sec.\,\ref{S_corona_ampp}), the AMPP automatically generates models ready to be populated with EBTEL solutions by computing the average magnetic field $\langle B\rangle$ and length $L$ of the magnetic field lines crossing each volume voxel.These parameters are used to compute the time-averaged volumetric heating rate, $\langle Q \rangle $, obtained from
\begin{equation}
\label{Eq_heating_rate_scaling}
 Q(t)  = Q_0 \left(\frac{\langle B\rangle}{B_0}\right)^a \left(\frac{L_0}{L}\right)^b f(t), 		\end{equation}
 and assign it to each voxel crossed by a closed field line. Here the heating profile $f(t)$ incorporates the duration $\Delta t$ of the nanoflares, the time interval between successive events $\tau$, and it may also include a dependence on mass density $\rho$. We have adopted the normalization convention $\langle f(t) \rangle \equiv1$, which, for any heating model, ensures that the averaged heating rate, $\langle Q\rangle$, stays the same for a fixed choice of the $Q_0$, $a$, and $b$ parameters.

There are five independent parameters: $Q_0$, $a$, $b$, $\tau$, and $\Delta t$. $Q_0$ is a typical heating rate, which can depend on the driver velocity $v$ and the electric current density along the flux tube, or, equivalently, on the force-free parameter $\alpha$, and so it can be different for different flux tubes. The actual numerical value of $Q_0$ (measured in erg cm$^{-3}$ s$^{-1}$) depends on the normalization constants, which are  chosen as $B_0=100\text{~G}$ and $L_0=2\times10^9\text{~cm}$. The power-law indices, $a$ and $b$, have certain values for a given heating model; for example $a=2$ and $b=1$ within the \textit{critical shear angle model} \citep{Mandrini2000}.

The time constants, $\tau$ and $\Delta t$, are additional free parameters of the model, which are informed by analysis of the EUV AR lightcurves \citep{Viall2012} and EBTEL modeling of these line-of-sight-integrated light curves \citep{Viall2013}. EBTEL is capable of accurately simulating the entire range of $\tau$, from effectively ``steady," to fully ``impulsive" (see \citet{Nita2018} for more details).

As detailed in \S\ref{section:DDM-DEM}, for a given set of input free parameters, the most recent version of the hydrodynamic simulation code, dubbed EBTEL++, outputs a pair of distributions over a relevant temperature range, which are the commonly used Differential Emission Measure (DEM):

\begin{equation}
\label{Eq_dem_def}
 \xi(T)=  \frac{n_e^2(T)dV}{VdT},  \quad {\rm [ cm^{-6}\, K^{-1}]}
\end{equation}

and the Differential Density Metrics \citep[DDM, ][]{2021ApJ...914...52F}%

\begin{equation}
\label{Eq_ddm_def}
 \nu(T)=  \frac{n_e(T)dV}{VdT},  \quad {\rm [ cm^{-3}\, K^{-1}]}.
\end{equation}

When only the DEM distributions are available, as was the case of the output provided by the original EBTEL code,  or if explicitly requested by the user, \gx uses them to assign effective density and temperature pairs to any model voxel crossed by a closed magnetic field line characterized by a \{$\langle B\rangle$, $L$\} pair:

\begin{eqnarray}
\label{n-T}
n_\xi&\equiv &\sqrt{\langle n_e^2\rangle}=\left(\int\xi(T)\,\mathrm{d}T\right)^{1/2},\\\nonumber
T_\xi&=&\frac{\int T\cdot\xi(T)\,\mathrm{d}T}{n_\xi}^2.
\end{eqnarray}

However, if the DDM distributions defined by Equation \ref{Eq_ddm_def} are also available,  as is the case of the output provided by the upgraded EBTEL++ code, \gx computes, by default, the effective density--temperature pairs defined as

\begin{eqnarray}
\label{n-T-DDM}   n_{\nu}&=&\int\nu(T)\,\mathrm{d}T,\\\nonumber
   T_{\nu}&=&\frac{\int T\cdot\nu(T)\,\mathrm{d}T}{n_{\nu}}.
\end{eqnarray}

Given the fact that a typical \gx model contains a large number of coronal voxels that need to be populated at the run-time with EBTEL solutions,  the practical approach that has been adopted is to run off-line the EBTEL code, to pre-compute several thousand combinations of the flux tube lengths, $L$, and nanoflare magnitudes (heating-model-specific time averaged volumetric heating rates), $\langle Q \rangle$, to create lookup tables that contain the coronal and transition region DEM and DDM distributions for each pair of flux tube length and nanoflare magnitude. Thus, using the $\langle B\rangle$ and $L$ properties computed by the pipeline for each coronal or transition region voxel,  the adjustable Equation \ref{Eq_heating_rate_scaling} is used at the run-time to select the corresponding nanoflare magnitudes and assign  the DEM and DDM distributions from pre-computed lookup tables to a given voxel, an approach that has been tested and validated by \citet{Nita2018}. By default, \gx assigns the DEM-DDM pair corresponding the closest \{$\langle Q\rangle, L$\} neighbor grid node found in the lookup tables, but the GUI provides a series of alternative irregular grid interpolation methods from which to choose, including 4-closest-neighbor weighted interpolation. 

The DEM distributions are used by the \gx EUV radiation transfer codes to compute synthetic emission maps corresponding to the SDO/AIA channels.  The effective thermal plasma density and temperature pairs inferred from the DDM or DEM distributions are used by the \gx GUI for volume  visualisation purposes, and by the legacy Fast Codes microwave rendering routines to  compute synthetic multi-frequency radio emission. However, following the recent upgrade of the microwave Fast Codes \citep{2021ApJ...914...52F}, \gx offers the option to select a rendering routine that employs these codes to compute synthesized microwave emission maps directly from the DEM/DDM distributions, as detailed in \S\ref{multi-thermal}.

\section{Customization of Active Region Pipeline Models}
\subsection{Command line customization scripts}
The current release of the \gx package includes a series of macro commands that allow batch mode customization of the pipeline models and generation of the multi wavelength synthetic maps, as well as a series of benchmark tools that may be used to perform quantitative model to data spectral and image comparison for the purpose of model validation, as described in this section.
To illustrate how  some of the macro commands included in the \gx package may be used to customize an AR model generated by AMPP and synthesize microwave emission maps corresponding to a user's selected field of view (FOV), we list in Appendix \ref{appendix:b} an IDL script that performs the following actions:
\begin{itemize}
    \item imports a magneto-thermal structure prepared by the AMPP script provided in Appendix \ref{appendix:a}.
    \item defines the desired FOV and spatial resolution for producing the synthesized maps.
    \item defines a set of volumetric heating rate parameters
    \item defines a heating rate formula following  Equation \ref{Eq_heating_rate_scaling},  which takes into account the user defined input parameters and the AMPP pre-computed $\langle B\rangle$ and $L$ voxel properties.
    \item selects a specific EBTEL table
    \item defines a set of frequencies for which the synthetic microwave maps will be computed
    \item calls a microwave rendering module that performs the geometrical rendering of the 3D models and solves the radiation transfer equation along each image LOS to produce the set of requested microwave maps
    \item saves on disk the output data produced by this script in two alternative forms: an SSW-compatible IDL map object and a GX-specific IDL structure that contains both image data, as well as metadata documenting the entire process involved in generating the script output.
\end{itemize}

\begin{figure}[!htb]
\centering
\includegraphics[width=0.7\columnwidth]{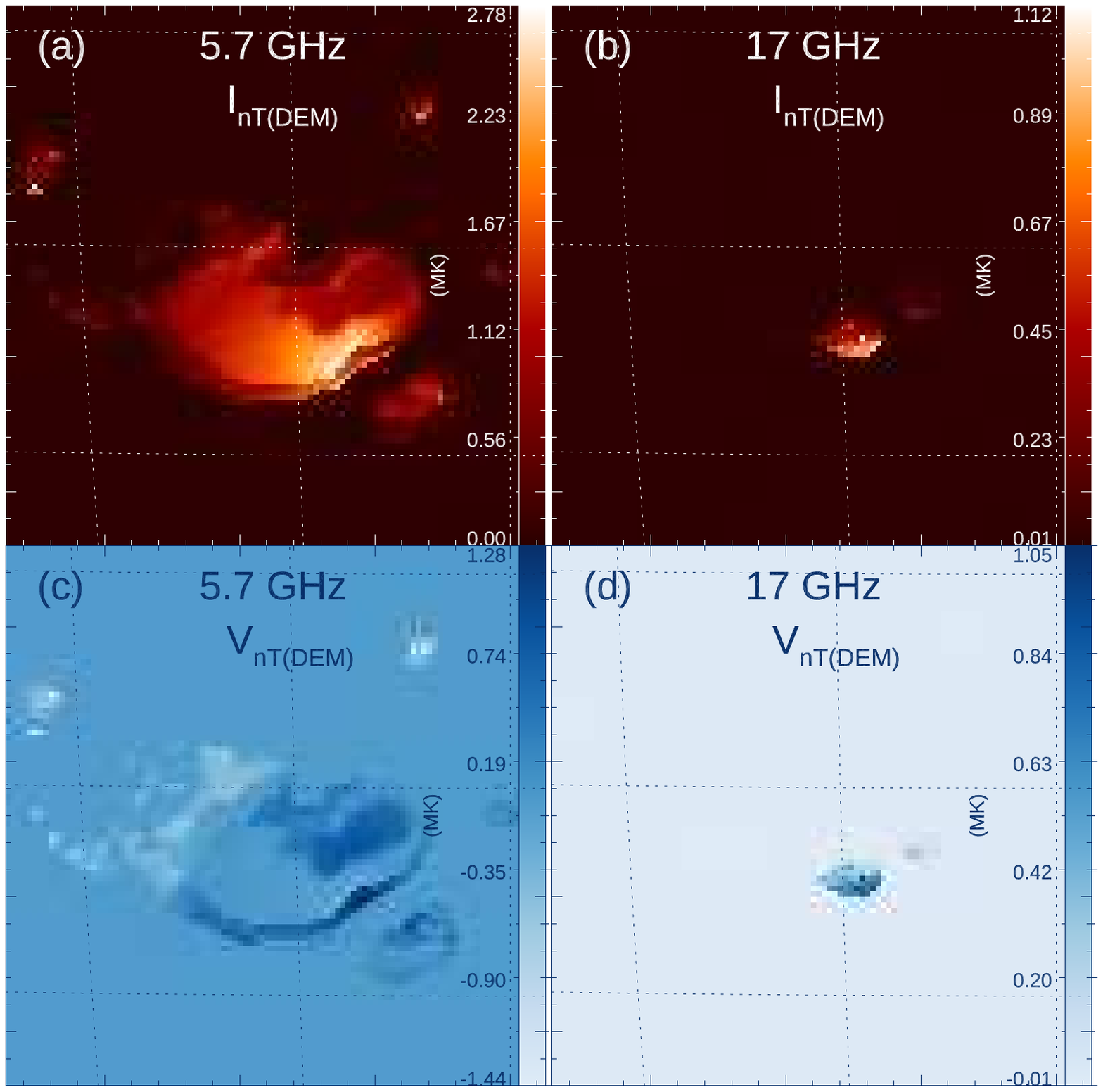}
\caption{ Brightness temperature Stokes $I$ (top row) and $V$ (bottom row) at 5.7 GHz (left column) and 17 GHz (right column) produced by the IDL script listed in Appendix \ref{appendix:b}.
\label{script_maps}}
\end{figure}

The Stokes $I$ and $V$ brightness temperature maps generated by this script are illustrated in Figure \ref{script_maps}. When combined with the data-to-model comparison tools described in \S\ref{model2data}, the script presented in Appendix \ref{appendix:b} may be employed to perform a fully automatic systematic search in a multidimensional parameter space for the combination of such magneto-thermal model parameters that produce synthesized maps that are simultaneously in best agreement with all multi-wavelength observational data available for a given instance of an AR, a methodology successfully employed by \citet{Fleishman11520} for finding the best scaling parameters within the low frequency heating assumption for the same instance of AR 11520 illustrated here.  
\vspace{0.5cm}
\subsection{Model to Data Comparison Tools
\label{model2data}}
The current release of GX\_Simulator provides a series of data-to-model comparison routines, located in the {\it{metrics}} sub-module, that are integrated in the GUI interface, but may be also called programmatically by customized analysis IDL scripts. The top level routine included in this submodule,  {\it{gx\_metrics\_map.pro}}, takes, as the input arguments, a model map structure and a reference map structure to be compared with, and returns a structure containing spatially resolved and FOV-averaged metrics such as absolute and normalized residuals, as well as $\chi^2$ metrics, if a map of standard deviations representing the observational or model uncertainties is provided as optional input.

By default, before computing the data to model metrics, this top level routine  also performs a map alignment by computing a cross correlation of the input map images, to which an optional mask may be applied. However, the user may choose not to align the input maps, or to apply a user-supplied spatial shift.

The FOV-averaged absolute and normalized residuals metrics returned by the top level routine are defined as follows:
\begin{eqnarray}
\label{res}
\langle{\cR}\rangle &=& \frac{1}{N}\sum_{ROI}^{}
\left({m_{ij}\otimes d_{PSF}}-{d_{ij}}\right),
\\\nonumber
\langle\rho\rangle &=& \frac{1}{N}\sum_{ROI}^{} \left( \frac{m_{ij}\otimes d_{PSF}}{d_{ij}}-1\right),
\\\nonumber
\langle\chi\rangle &=& \frac{1}{N}\sum_{ROI}^{} \left( \frac{m_{ij}\otimes d_{PSF}-d_{ij}}{\sigma_{ij}}\right),
\end{eqnarray}
where $d_{ij}$ is the observed brightness in the image pixel $_{ij}$ and  $m_{ij}\otimes d_{PFS}$ is the corresponding model map brightness convolved with the instrumental point spread function (PSF), the model or observational uncertainties are denoted by $\sigma_{ij}$, and $N$ represents the total number of pixels in the region of interest (ROI) over which the comparison is made (which may be all or only a subset of the FOV pixels selected by applying an optional, user defined byte mask).

\begin{figure}[!htb]
\centering
\includegraphics[width=0.8\columnwidth]{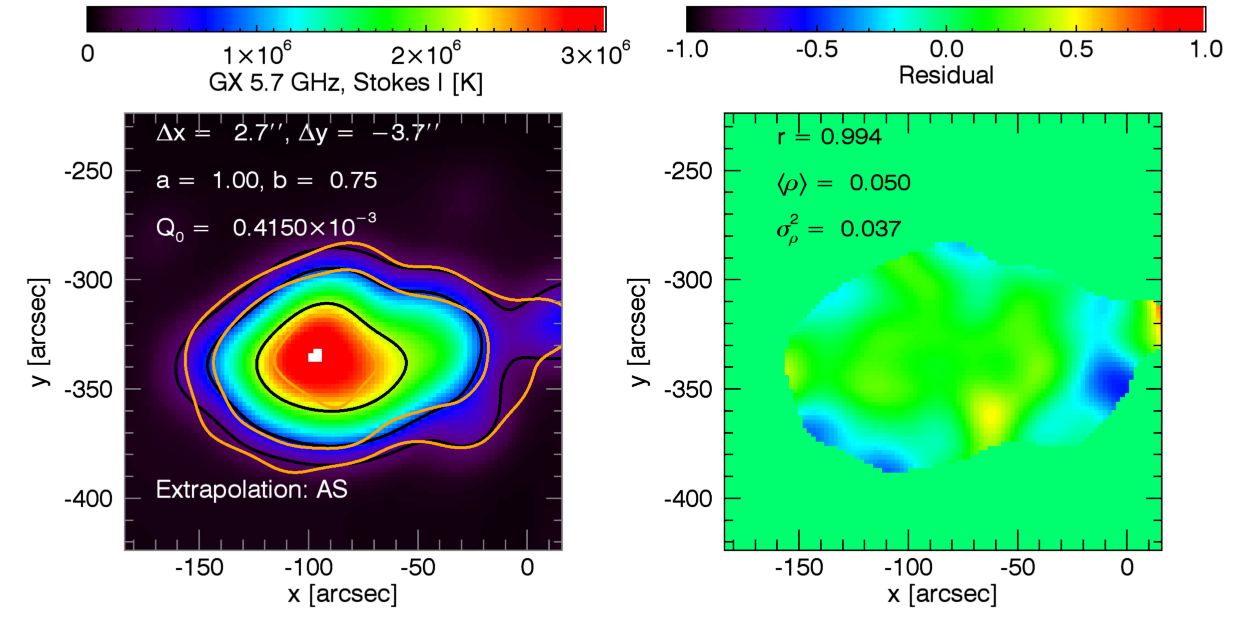}\\
\includegraphics[width=0.8\columnwidth]{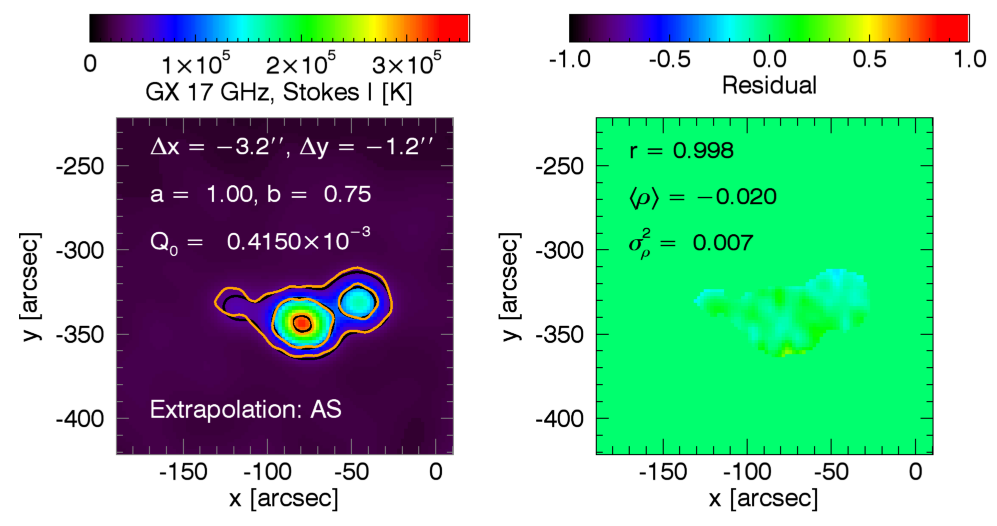}
\caption{
\label{fig_metrics}
Model to data comparison for the AR11520 model (reproduced from \citet{Fleishman11520}). Top row-left panel: The [12, 30, 80]\% contours of the 5.7 GHz SSRT observational map (yellow) and the synthetic map (black) are shown on top of synthetic 5.7 GHz map background image. The alignment shifts, $\Delta x$ and $\Delta y$, and the heating parameters, $Q_0$, $a$, and $b$ are indicated in the figure inset. Top row-right panel: corresponding model to data residual map normalized by the observed SSRT brightness temperature.  Pearson  cross-correlation coefficient, $r$, the averaged normalized residual, $\rho$, and the squared residual metrics, $\rho^2$, are indicated in the figure inset. Bottom row: the same data to model comparison as in the top row, between the 17~GHz synthetic image and the NoRH 17~GHz image.}
\end{figure}

The corresponding squared metrics are defined as follows:
\begin{eqnarray}
\label{FOV_res}
\langle{\cR}^2\rangle &=& \frac{1}{N}\sum_{ROI}^{}
\left( {m_{ij}\otimes d_{PSF}}-{d_{ij}}\right)^2.
\\\nonumber
\langle\rho^2\rangle &=& \frac{1}{N}\sum_{ROI}^{}
\left( \frac{m_{ij}\otimes d_{PSF}}{d_{ij}}-1\right)^2 ,
\end{eqnarray}
and the metrics of success for the data to model comparison are defined as
\begin{eqnarray}
\label{res2}
\sigma_\rho^2 &=& \frac{1}{N}\sum_{ROI}^{}
\left( \frac{m_{ij}\otimes d_{PSF}}{d_{ij}}-1-\langle\rho\rangle\right)^2 = \langle\rho^2\rangle-\langle\rho\rangle^2
\\\nonumber
\sigma_{\cR}^2 &=& \frac{1}{N}\sum_{ROI}^{}
\left({m_{ij}\otimes d_{PSF}}-{d_{ij}}- \langle{\cR}\rangle\right)^2=
\langle{\cR}^2\rangle - \langle{\cR}\rangle^2,
\\\nonumber
\langle\chi^2\rangle &=& \frac{1}{N}\sum_{ROI}^{}
\left( \frac{m_{ij}\otimes d_{PSF}-d_{ij}}{\sigma_i}\right)^2- \langle{\chi}\rangle^2,
\end{eqnarray}
where, as motivated by \citet{Fleishman11520}, the last term of each of the metrics of success defined above is subtracted to account for any imperfections in fine tuning the model, which may result in minimized but not exactly null averaged residuals that, as defined by Equation \ref{res}, are expected to vanish in the case of a perfect data to model match. As an example of data to model comparison performed using the \gx metrics routine, we reproduce in  Figure \ref{fig_metrics} the results obtained by \citet{Fleishman11520} using observational maps obtained with data from the Siberian Solar Radio Telescope \citep[SSRT,][]{ssrt} and the Nobeyama Radio Heliograph \citep[NoRH,][]{norh} and the  5.7~GHz and 17~GHz synthetic images shown in Figure \ref{script_maps}(a,~b), which were convolved with the corresponding instrumental beams before their coalignment with the observational maps.

\subsubsection{Coronal Heating Modeling Pipeline (CHMP)}

To identify the combination of $\{a,b\}$ and $Q_0$ parameters that provides the best possible model to data agreement, quantitatively measured by the metrics described in \S\ref{model2data},
we have included in the most recent release of the \gx package a macro routine, namely {\it\textbf{gx\_search4bestq.pro}}, which, starting from a pair of initial guess heating rates, $\{Q_{0_{1}},Q_{0_{2}}\}$, performs a self-adaptive search for the optimal EBTEL heating rate $Q_0$ corresponding to a predefined $\{a,b\}$ pair chosen by the user.

The {\it\textbf{gx\_search4bestq.pro}} macro routine provides the core functionality for a top-level command line application, namely {\it\textbf{chmp.pro}}, which allows the user to interactively setup and start a multi-threaded search for the best possible EBTEL model over $\{a,b\}$ parameter grid, which takes advantage of the fact that such search is an embarrassingly parallel process.

\begin{figure}[!htb]
\centering
\includegraphics[width=0.95\columnwidth]{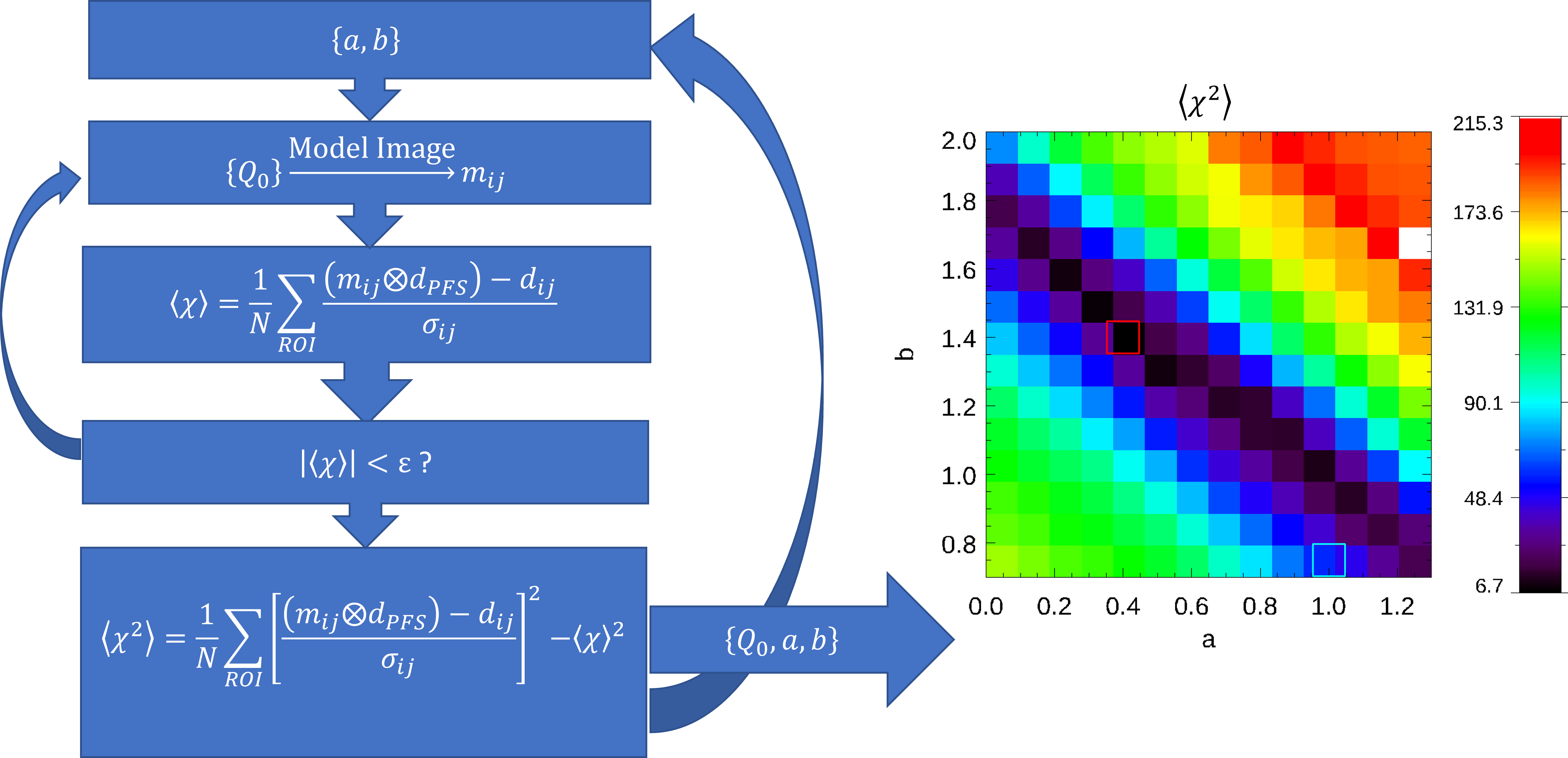}
\caption{ \label{fig_chmp}
The automated CHMP architecture and illustrative example. Left panel: The CHMP iterative flowchart. The outer loop steps through a user-defined grid of $\{a,b\}$ parameter pairs for which the inner loop employs the {\it\textbf{gx\_search4bestq.pro}} core macro to search for the optimal heating rate $Q_0$ by attempting to bring the absolute value of $\langle{\chi^2}\rangle$ metrics below a predefined maximum threshold $\epsilon$. Right panel: The output of the automated CHMP search using NoRH 17~GHz observational data for the same AR11520 \gx model as previously tuned by \citet{Fleishman11520}. The image displays, in logarithmic scale, the $\langle{\chi^2}\rangle$ corresponding to each investigated grid point. As indicated by the right side color bar, the $\langle{\chi^2}\rangle$ metrics range in this case between 6.70 and 215.3. The red rectangle marks the grid point corresponding to the best found metrics $(a=0.4,b=1.4; \langle{\chi}\rangle^2=6.70)$, which may be compared with the $\langle{\chi^2}\rangle=13.98$ metrics corresponding to the best parameter combination, $(a=1.0,b=0.75)$, found by \citet{Fleishman11520} by minimizing the $\sigma^2$ metrics.
}
\end{figure}

The options provided by the CHMP command line application may be explored by calling the routine with no keyword arguments, which generates the following console messages:

\begin{verbatim}
\input{chmp_cmd.bat}
\end{verbatim}
The left panel of Figure \ref{fig_chmp} displays a flowchart  that illustrates the iterative search process of the CHMP application. As an illustrative example, the right panel of Figure \ref{fig_chmp} displays the distribution of the best  $\langle{\chi^2}\rangle$ metrics over the searched parameter space that has been obtained for the same AR11520 \gx model that was previously tuned by \citet{Fleishman11520} using a direct trial and error approach, which sought to minimize the $\sigma_\rho^2$ metrics using reference observational microwave data at 17~GHz provided by NoRH.

The optimal parameter combination found by the automated CHMP application in this illustrative example, $(a=0.4,b=1.4, \langle{\chi}\rangle^2=6.70)$, is marked as red squared on the metrics image shown in the right panel of Figure \ref{fig_chmp}. It is different from the brute-force solution found by \citet{Fleishman11520},  $(a=1.0,b=0.75, \langle{\chi}\rangle^2=13.98)$, marked as a cyan rectangle on the grid, although the data-to-model comparison metrics are of the same order of magnitude. As revealed by the $\langle{\chi}^2\rangle$ metrics distribution image, both solutions belong to the same region of comparatively good metrics, which stand out as a diagonal path against the surrounding parameter space.
We interpret this preliminary result as an indication of a certain degree of degeneracy of the EBTEL solution, which we speculate might be possible to remove by combining the results of a CHMP search independently performed at more than one observational frequency, an avenue that we consider worth being pursued, but out of the scope of this paper.

\subsubsection{CHMP graphical user interface (GX\_CHMP)}
\begin{figure}[!htb]
\centering
\includegraphics[width=0.75\columnwidth]{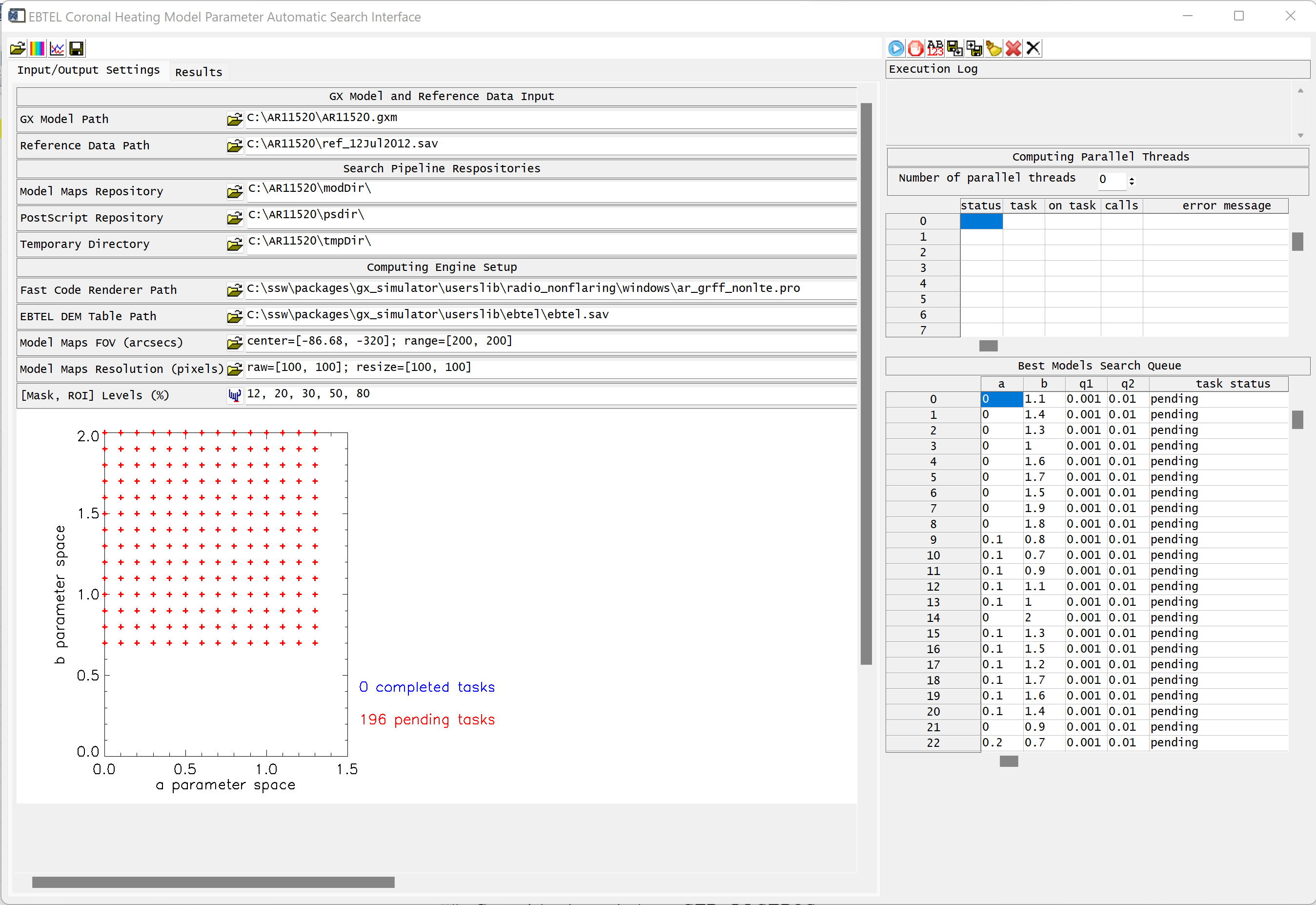}\textbf{(a)}\\
\includegraphics[width=0.75\columnwidth]{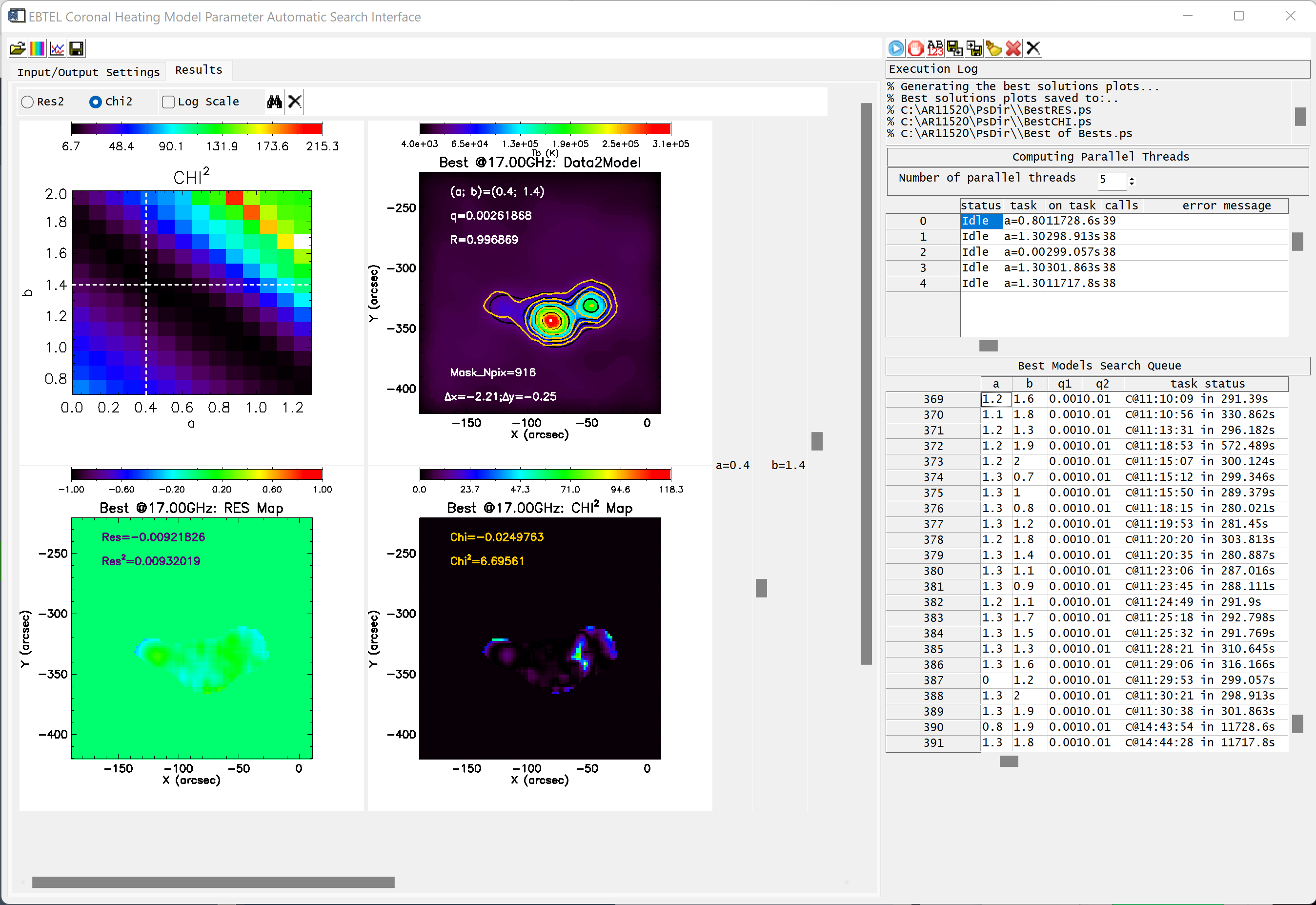}\textbf{(b)}
\caption{ \label{fig_gxchmp}\gx CHMP Graphical User Interface (GX\_CHMP): Panel (a): Search setup input fields tab. Panel (b): Output solution display area tab: top left plot: Best $\langle\chi^2\rangle$ (or $\sigma^2_\rho$) metrics corresponding to each grid point. The intersection of dotted vertical and horizontal lines indicates a grid point selected by the user, which, in this case,  corresponds  to the best metrics found. Top right plot: model-convolved map (background) and model (black) and data (yellow) user defined contours delimiting the ROI area; bottom right plot: map of the $\sigma^2_\rho$ metrics corresponding to the selected grid point in the top left panel; bottom left plot: map of the $\chi^2$ metrics corresponding the selected grid point in the top left pane. The regions outside the ROI area are set to neutral uniform values in both metrics maps.}
\end{figure}

In addition to the CHMP command line application, the core {\it\textbf{gx\_search4bestq.pro}} routine includes many more built in options that provide more user flexibility, including the option of performing a search over an arbitrarily shaped, or sparse, grid space. The full flexibility of the {\it\textbf{gx\_search4bestq.pro}} routine is exposed to the user by the {\it\textbf{gx\_chmp.pro}} application. Figure \ref{fig_gxchmp} shows this GUI application, which allows one to take advantage of all available options provided by the core macro routine. It also interactively visualizes, at run-time, the data to model comparison metrics maps corresponding to each grid point for which a solution has been computed.

\vspace{12pt}
\subsection{DEM and DDM EBTEL/EBTEL++ Tables}
\label{section:DDM-DEM}
\gx includes EBTEL results from a number of different coronal heating scenarios. Each is provided in the form of an IDL {\it sav} file containing the following floating point array variables:
\begin{itemize}
\item LOGTDEM$[N_T]$: logarithm of temperature bins over which the DEM/DDM distributions are computed
\item QRUN$[N_Q \times N_L]$: the average heating rates, $\langle Q\rangle$ corresponding to each grid point
\item LRUN$[N_Q \times N_L]$: the loop half lengths, $L_{1/2}\equiv L/2$, corresponding to each grid point
\item DEM\_COR\_RUN$[N_T\times N_Q \times N_L]$: DEM distributions for coronal voxels
\item DEM\_TR\_RUN$[N_T\times N_Q \times N_L]$: DEM distributions for transition region voxels
\item DDM\_COR\_RUN$[N_T\times N_Q \times N_L]$:  DDM distributions for coronal voxels
\item DDM\_TR\_RUN$[N_T\times N_Q \times N_L]$: DDM distributions for transition region voxels
\item TRUN$[N_Q \times N_L]$:  maximum electron temperature achieved at any point during the associated EBTEL run (only included for informational purposes, but not used by \gx)
\end{itemize}
 where $N_T$ denotes the number of temperature bins in the DEM/DDM distributions, and ${N_Q\times N_L}$ represents the dimension of the parameter grid space over which the distributions have bin computed.

The same as the previous versions of \gx, the current version includes two EBTEL tables representing impulsive and steady heating. The impulsive heating table was generated using triangular nanoflare profiles with  $\Delta t=20$~s and $\tau=10,000$~s, while the steady heating table was generated with constant heating. These two tables cover the same $\sim10^6~\text{cm} \leq L_{1/2} \leq\sim4\times10^{10}~\text{cm}$ range, but have different time averaged heating rate, $\langle Q\rangle$, ranges, with the steady heating encompassing  $\sim3\times10^{-9}~\text{erg}~\text{cm}^{-3}~\text{s}^{-1} \leq \langle Q\rangle ~\leq\sim2\times10^{6}~\text{erg}~\text{cm}^{-3}~\text{s}^{-1}$ and the impulsive heating covering $\sim 6\times10^{-10}~\text{erg}~\text{cm}^{-3}~\text{s}^{-1} \leq \langle Q\rangle ~\leq\sim0.2~\text{erg}~\text{cm}^{-3}~\text{s}^{-1}$ on different irregular grids.

In addition to these idealized heating scenarios, the current version of \gx also includes six new tables designed to emulate more physically realistic coronal heating conditions. Each of the EBTEL++ models \citep[][code available on GitHub\footnote{\href{https://github.com/rice-solar-physics/ebtelPlusPlus}{https://github.com/rice-solar-physics/ebtelPlusPlus}}]{Barnes2016a} used to generate these new tables includes a range of heating event sizes and time delays between successive events combined with steady background heating. This produces stochastic heating more representative of the true conditions on the Sun. These grids cover a standard parameter space of $10^6~\text{cm} \leq L_{1/2}\leq \sim6\times10^{10}~\text{cm}$ and $10^3~\text{erg}~\text{cm}^{-2}~\text{s}^{-1} \leq \langle Q\rangle L_{1/2} \leq \sim8\times10^8~\text{erg}~\text{cm}^{-2}~\text{s}^{-1}$, sampled with a resolution of 0.1 dex for a total of 2940 models. This construction of the heating rate ensures that all models encompass observed heat flux ranges \citep{Withbroe1977} and is largely consistent with the parameter spaces covered by the original two tables.

In these new EBTEL++ tables the steady heating term is chosen to sustain loops with apex temperatures of $\sim0.3~\text{MK}$, well below typical coronal temperatures. This is achieved by applying a steady background heating of $6.3\times10^{12}/L_{1/2}^2 \geq 10^{-7}~\text{erg}~\text{cm}^{-3}~\text{s}^{-1}$ where the minimum threshold only impacts the longest loops, which can become unstable without it. In cases where this background heating term equals or exceeds $\langle Q\rangle$, it is capped at $\langle Q\rangle$, resulting in cooler loops with no impulsive heating. As a consequence of this steady heating term, a significant fraction of the models in the new EBTEL++ tables (those with lower heat flux, particularly for the shortest and longest loops) undergo identical steady heating in each table. However, since this occurs in the least physically realistic corners of the parameter space it should have only a minor impact on the resulting \gx models.

The impulsive heating in the new EBTEL++ tables is their only differentiator. Each heating event is a triangular heat pulse with a $\Delta t=100$~s duration \citep[for a discussion of why longer duration nanoflares may produce more realistic results see][]{Barnes2016a} with an amplitude drawn from a power law distribution of event sizes. The amplitude of each heating event is correlated with the delay until the next heating event, with a proportionality (accounting for the steady background heating) that enforces $\langle Q\rangle$ from the start of the heating event until the start of the next event. In practice, these amplitudes are computed from the time delay that is drawn randomly from a power law probability distribution defined by the slope $\alpha$, the median time between heating events $\langle\tau\rangle$, a maximum time between events chosen as $3\langle\tau\rangle$, and the automatically determined minimum delay. The median time between heating events is defined as some scaling multiple of the cooling time $\tau_0$ which depends on $\langle Q\rangle$ and $L_{1/2}$, \citep{Cargill2014, Barnes2019} and the ratio of the longest delay (corresponding to the strongest event) to $\langle\tau\rangle$, in this case 3. The six new tables included in \gx are computed with each combination of $\alpha=-1$ (large-) or $-2.5$ (small-event dominated) and $\langle\tau\rangle=0.2\tau_0$ (high-), $1\tau_0$ (intermediate-), or $5\tau_0$ (low-frequency heating). All the EBTEL/EBTEL++ tables distributed with the current version of \gx along with their $\langle\tau\rangle$ and $\alpha$ parameters are listed in Table \ref{tab:ebtel_files}.

\begin{table}[h!]
  \begin{center}
    \caption{EBTEL coronal heating tables included in the \gx distribution package}
    \label{tab:ebtel_files}
    \begin{tabular}{l|c|c}
    \hline
    \hline
      \textbf{Filename} & \textbf{$\langle\tau\rangle$} & \textbf{$\alpha$} \\
      \hline
        ebtel.sav                                &   $\tau=10^4$s    &   N/A     \\
        \text{ebtel\_ss.sav}                    &   constant        &   N/A     \\
        \text{ebtel\_scale=0.2\_alpha=-1.sav}    &   0.2$\tau_0$     &   -1      \\
        \text{ebtel\_scale=0.2\_alpha=-2.5.sav}  &   0.2$\tau_0$     &   -2.5    \\
        \text{ebtel\_scale=1\_alpha=-1.sav}      &   $\tau_0$        &   -1      \\
        \text{ebtel\_scale=1\_alpha=-2.5.sav}    &   $\tau_0$        &   -2.5    \\
        \text{ebtel\_scale=5\_alpha=-1.sav}      &   5$\tau_0$       &   -1      \\
        ebtel\_scale=5\_alpha=-2.5.sav    &   5$\tau_0$       &   -2.5    \\
    \hline
    \end{tabular}
  \end{center}
\end{table}

For each \{$\langle Q\rangle$, $L_{1/2}$\} pair in the new EBTEL++ tables, the model is run for $1000\langle\tau\rangle$ and the heating events are drawn using the same random seed, i.e. the heating events are identical except for changes in the $\langle\tau\rangle$ and $\langle Q\rangle$ scaling. The coronal and transition region DEMs and DDMs for the model are generated by averaging the time-evolving DEMs and DDMs over the entire model run, excluding the first $10\langle\tau\rangle$ to ensure that the model's initial conditions (static equilibrium at the background heating rate) do not influence the results. The same time-averaging technique is applied in the original EBTEL tables except the DEMs and DDMs are averaged over the entire model run (the original EBTEL runs assumed either steady heating or homogeneous nanoflares with a fixed delay of 10,000 s). This time averaging simulates the behavior of many small magnetic strands experiencing heating with the same properties but out of phase. Since the corona is typically optically thin and individual magnetic strands have cross sections well below observable resolutions, this is a simple way to represent the observable plasma distributions. More details about similar EBTEL++ models can be found in \cite{Schonfeld2020} and the code used to generate these EBTEL++ tables is available from a Zenodo\copyright\ digital library \citep[][ doi.org/10.5281/zenodo.7154827]{ebtel_zenodo}.

\begin{figure}[!htb]
\centering
\includegraphics[width=0.8\columnwidth]{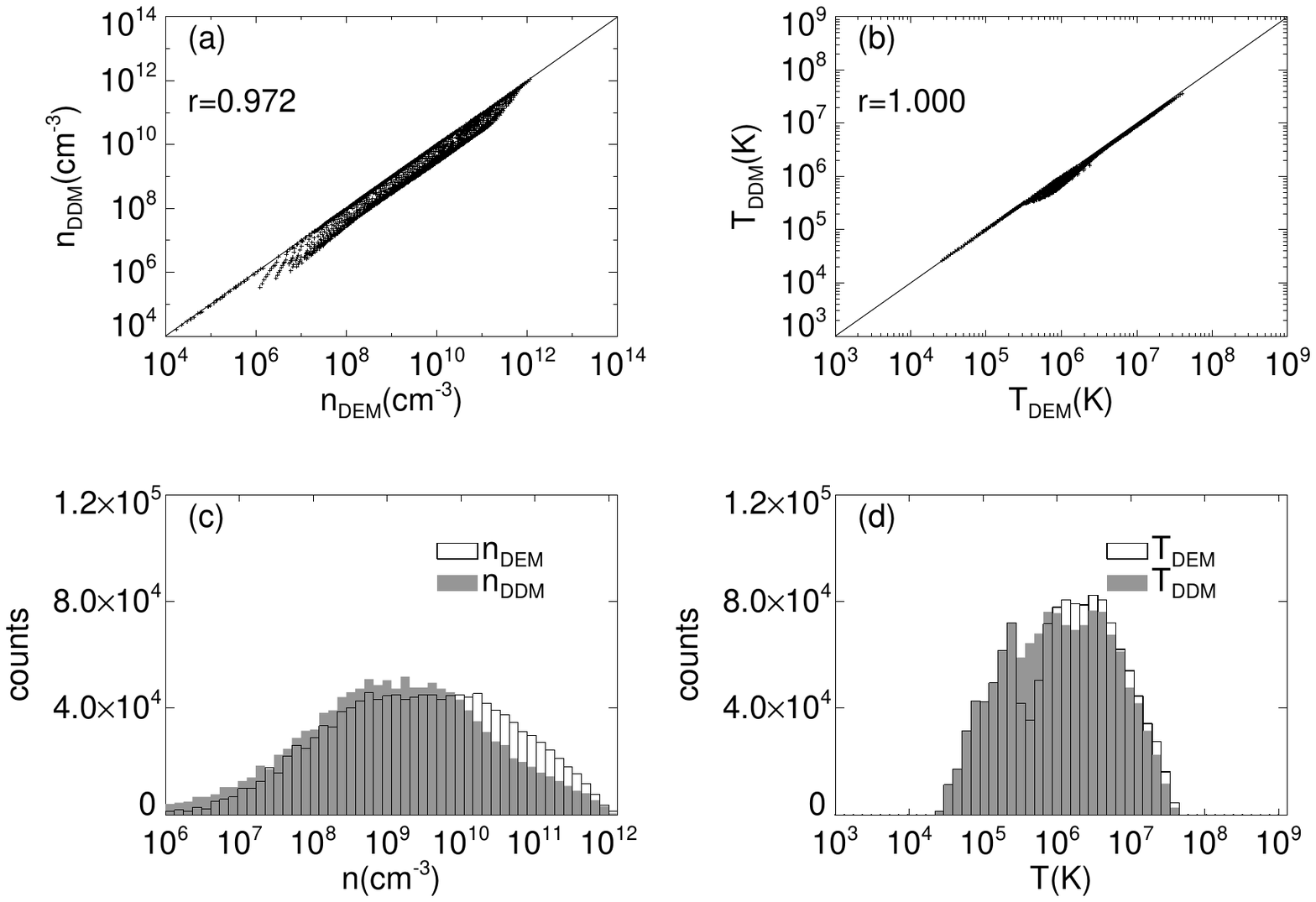}
\caption{
\label{ddm-dem}
Comparison of thermal electron densities and temperatures pairs computed from the moments of the DDM distributions (Equation \ref{n-T-DDM}) versus those computed from moments of the DEM distributions (Equation \ref{n-T}) for the {ebtel\_scale=5\_alpha=-2.5.sav} table. Top row panels: $n_{DDM}$ vs. $n_{DEM}$ (panel a) and $T_{DDM}$ vs. $T_{DEM}$  (panel b) correlation plots indicating a good linear correlation quantified by the correlation coefficients indicated in each panel. Bottom row panels: comparison between the DEM-derived distributions (black histograms) and the corresponding  DDM-derived distribution (grey filled histograms) indicating a systematic shift towards lower values of the latter.}
\end{figure}

As described in \S\ref{section:corona}, \gx uses Equation \ref{n-T-DDM}, when DDM distributions are available, or Equation \ref{n-T} if only DEM distributions are available, to assign effective density--temperature pairs to each voxel crossed by a magnetic field line characterize by the averaged magnetic field $\langle B\rangle$ and length $L$ for the purposes of visualization, or for computing emission using one of the radiation transfer codes that require them as input parameters. To help assess the differences between the parameters computed using these two approaches, we present a direct comparison in Figure \ref{ddm-dem}.

The top row panels in Figure \ref{ddm-dem} show the density (panel a) and temperature (panel b) correlation plots between the respective parameters computed as moments of the DEM distributions (Equation \ref{n-T}) and the DDM (Equation \ref{n-T-DDM}) distributions provided by one of the EBTEL++ tables distributed with the \gx package, namely \textbf{ebtel\_scale=5\_alpha=-2.5.sav}, which were pre-computed for a set of 2,940 pairs of averaged volumetric heating rates, $\langle Q\rangle$ , and magnetic loop half lengths, $L_{1/2}$, spanning a parameter space ranging between $( 1.58\times10^{-8}-794.32)$ erg cm$^{-3}$ s$^{-1}$ and $(10^6-6.31\times10^{10})$ cm, respectively. In addition, the bottom row panels in Figure \ref{ddm-dem} show a comparison between the corresponding density (panel c) and temperature (panel d) distributions computed by these two alternative means. Figure \ref{ddm-dem} demonstrates the expected good correlations of the DEM/DDM density and temperatures and a systematic shift towards lower values of the DDM-inferred $n$ and $T$ distributions.\\

\subsection{ Synthesized Microwave Emission from Multi-Thermal Plasma Models}\label{multi-thermal}
\begin{figure}[!htb]
\centering
\includegraphics[width=0.7\columnwidth]{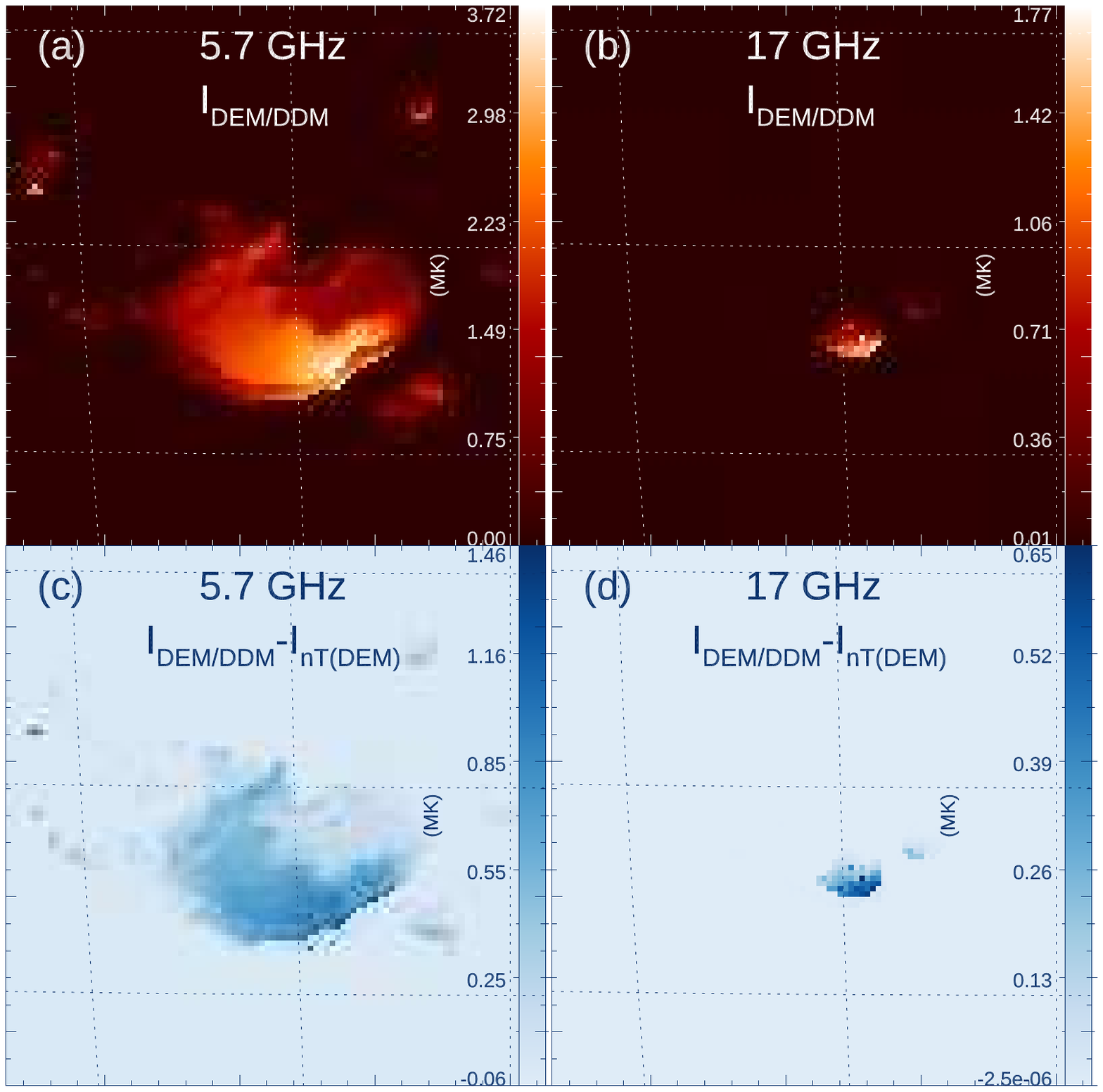}
\caption{Comparison between the brightness temperature maps for AR 11520 computed using the full DEM/DDM treatment and the isothermal  approximation, (shown on the top row in Figure \ref{script_maps}), at frequencies 5.7 GHz (left column) and 17 GHz (right column). Top row: maps based on the multithermal DEM/DDM treatment; bottom row: the respective difference maps. The color bars are in MK units.}
\label{DEMimageCompare}
\end{figure}

To illustrate the effect of the multi-thermal plasma distributions on the radio emission, we computed radio maps for the above-mentioned AR 11520 model, using the full DEM/DDM treatment \citep[][see also Section \ref{Sec_Rad_Trans}]{2021ApJ...914...52F}. The resulting 5.7 and 17 GHz brightness temperature maps are displayed in Figure \ref{DEMimageCompare}, where they are compared with the co responding maps displayed in Figure \ref{script_maps}, which were obtained using the  ``classical'' isothermal treatment (based on the DEM moments, according to Eqs. (\ref{n-T})).  The bright emission at these frequencies is produced mainly due to the thermal gyroresonance mechanism, which is sensitive to the DDM distribution, while the contribution of the thermal free-free emission (which is sensitive to the DEM distribution) is relatively low. One can see in Figure \ref{DEMimageCompare} that considering the multi-thermal plasma composition affects significantly both the image morphology (especially at lower frequencies) and the peak or average brightness temperatures; for the multi-thermal model, the brightness temperatures are higher due to contribution of electrons with energies above the average energies.

\subsection{Integration of User-Supplied Models in Pipeline-Produced Model Skeletons }
Although the top-level AMPP IDL procedure, i.e. {\bf gx\_fov2box.pro}, provides the user with a series of keyword switches to produce different types of standard  \gx models, for maximum flexibility the experienced user may choose to design a custom AMPP IDL script by combining the low-level AMPP routines provided by the \textbf{gxbox} sub-module. Moreover, the modular architecture of AMPP also allows for custom-designed computation blocks to be inserted to replace a standard block, at any point following the empty box creation step, provided that the IDL box structures produced by such customized blocks remain compatible with the \gx architecture. Such customized box structures may contain any number of additional, non conflicting tags, which would be quietly ignored by \gx.

For example, one may fill the empty-box \textbf{.NONE.} IDL structure produced by the AMPP initialization block with a magnetic field model obtained by alternate means. Similarly, one may replace an AMPP-generated chromo model with a non-standard one, provided that all chromosphere-related tags of a standard \textbf{``.CHR."} box structure are either replaced with equivalent data, or kept unaltered.

Although the current \gx release has no provisions for allowing an alternative approach to our EBTEL-based coronal model, one may generate a \gx compatible model populated with ready-to-use numerical thermal density and temperature pairs produced by any means (e.g. MHD simulations), as described in Appendix \ref{appendix:c}.

\section{Creation and Customization of Flaring Loop Models} \label{section:loops}
The core of the flaring loop modeling capabilities included in the previous versions of the \gx package have been described in \citet{Nita_etal_2015}. In addition, the current version of \gx includes the ability of using array-defined electron distributions for calculating the nonthermal radio emission, and a scripting feature for automation that allows users to step through many simulations to follow the temporal evolution of a flare.

The workflow of creating a flare model is to begin with a magnetic field model, usually generated by AMPP as described in \S\ref{pipeline}, identify a likely field line in the magnetic field model that corresponds to the core, or axis, of a flaring loop, populate that loop with nonthermal particles (embedded in a non-flaring background solar atmosphere), and then calculate the emission of choice (among multiple EUV passbands, X-ray energies, or microwave frequencies). If one is interested in the initiation of a flare, a vector magnetogram close in time but prior to the flare may be the best choice, especially when a newly formed flux rope is involved.  To model emission from a loop that has formed as a result of a flare, a vector magnetogram taken sometime after the flare may be a better choice.
Once the flaring loop has been selected and populated, the properties of the model may be adjusted, and loops added, consistent with the spectroscopic imaging observations (i.e. parallel to EUV loops or joining HXR footpoints), until the emission from the model (when convolved with the instrument’s PSF adequately fits all observational constraints. Matching not only the morphology, but also the frequency/wavelength dependence (shape and brightness of the microwave spectra for example) is highly constraining and, when successful, the resulting forward fit model accounts naturally for the source inhomogeneity and finite instrument resolution limitations that adversely affect spectral inversions.

\begin{figure}[!htb]
\centering
\includegraphics[width=0.9\textwidth]{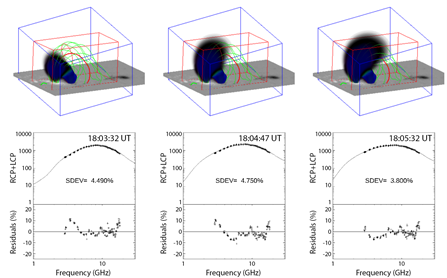}\\
\includegraphics[width=0.9\textwidth]{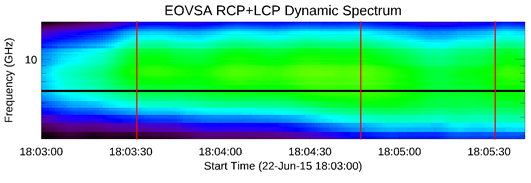}
\caption{\label{flare-evolution} Dynamic 3D modeling with GX Simulator for the peak of the SOL2015-06-22T17:50 flare observed with EOVSA. Top row: Analytically defined nonthermal electrons distributions in a system formed by four loops. Middle row: Model (lines) to EOVSA data (symbols) spectral comparisons and relative residuals, corresponding to the selected time frames illustrated in the top row. Bottom row: EOVSA Dynamic spectrum during the fitting interval (2.4-18 GHz, 4 seconds time resolution), with vertical red lines showing the times illustrated in the upper rows. }
\end{figure}

Once a successful model is obtained for one time in the flare, the new \gx’s scripting capability allows similar models to be quickly generated with evolving parameters to follow the temporal evolution of the flare. \citet{FlareEvolution2018} employed this sequential approach to model the evolution of the radio emission observed by EOVSA during the peak phase of the SOL2015-06-22T17:50 M6.5 solar flare. \citet{FlareEvolution2018} generated an evolving sequence of 30 models obtained by fine-tuning the analytically defined non-thermal electron distributions populating two of the four flux tubes that were previously identified and used by \citet{Kuroda2018} to model a single time frame corresponding to a local peak at 18:05:32 UT. Figure \ref{flare-evolution} displays three selected snapshots of this evolving model evolution (top row) and the corresponding match between the FOV-integrated synthetic microwave spectra computed by \gx and the corresponding observational spectra produced by EOVSA.

Until recently (including the studies by \citep{FlareEvolution2018} and \citet{Kuroda2018} mentioned above), the nonthermal particle distribution that could be assigned in \gx to a flaring loop model was limited to a single energy distribution. Thus, when a value for a powerlaw index and pitch angle were chosen for a flux tube, those same values were applied everywhere in the loop.  The fast codes that calculate microwave emission have now been upgraded to permit array-defined distributions in energy and pitch angle that can vary in an arbitrary manner \citep[][see \S\ref{fastcodes} for more details]{UltimateFastCodes2021}.  This upgrade opens the possibility to create a wide range of particle distributions as a function of time and position  along a loop, based for example on 1D Fokker-Planck simulations, and from them calculate the emission maps as before.  The importance of this approach to the particle transport problem is that a time-dependent, physics-based simulation with varying levels of enhanced energy and pitch-angle diffusion can now be explored in a realistic loop geometry and compared quantitatively with the actual multi-wavelength observational data provided by instruments such as EOVSA, SRH, RHESSI, or STIX.

\begin{figure}[!htb]
\centering
\includegraphics[width=0.9\textwidth]{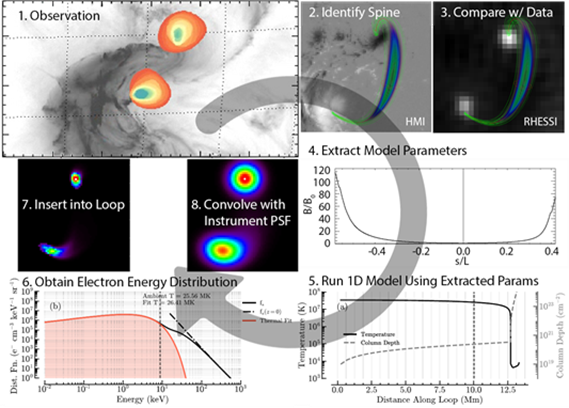}\
\caption{\label{flare-workflow} Illustration of the workflow from observation to simulated images. (1) The multifrequency EOVSA images for one time frame in the 2017 Sep 04 flare overlaid on an AIA 131 A image, showing two widely separated footpoints. (2) The HMI vector magnetogram (background gray scale is $B_z$) is used to calculate a NLFFF magnetic model from which a flux tube central field line is identified. (3) Other imaging data such as this comparison with RHESSI HXRs are used to refine the choice of central field line. (4) The model atmosphere parameters are extracted along the chosen field line. (5) The model parameters are used with a 1D Fokker-Planck code, starting from some assumed injected distribution and adjustable diffusion coefficients, to calculate (6) the energy distribution as a function of position and time.  The energy distributions along the loop are then used by \gx to calculate (7) the microwave emission maps, which are then (8) convolved with the EOVSA’s PSF for quantitative comparison with EOVSA data. }
\end{figure}

To make use of this added functionality, the current version of \gx allows the user to export the physical and geometrical properties (B, $n_e$, T, $\alpha$) along the axis of a selected flux tube. These properties may be then used in a time-dependent 1D particle transport code to calculate externally such numerical particle distributions. Then they can be imported back into the GX model and assigned to each volume element, from which various emissions can be calculated.  This workflow is illustrated in Figure \ref{flare-workflow} for the event of 2017 Sep 04, an M5.4 event.

Following the steps in the order indicated by the gray circular arrow, one starts with (1) the multifrequency observations (EOVSA in this example, multi-colored filled contours) and use them to identify (2) the magnetic field line that fits the morphology.  Then, (3) one may further compare with other data (RHESSI image in this example) to fine-tune the selection. Once a suitable axis of a flaring flux tube is identified matching the morphology of the flare, one has to quantitatively adjust plasma and particle parameters in the loop based on fitting all parameters inferred from observations (e.g. EOVSA, AIA, and other diagnostics).  Then, (4) one has to extract from the model the magnetic field, density, temperature, and other atmospheric parameters as a function of distance along the axis. Panel 4 of Figure \ref{flare-workflow} illustrates the value of $B/B_0$ for the particular loop chosen for illustration.
From here, any model that can calculate a 1D time-dependent electron distribution $f(E,\mu,s,t)$ can be used, where $E$ is the electron energy, $\mu$ is the cosine of pitch angle, $s$ is the distance in the 1D model, and $t$ is the time.

One such model framework is the 1D electron transport simulations along the loop, wherein a source term injection of particles is introduced and the evolution of that distribution as it diffuses in both pitch angle and energy  provides the required distribution along the loop.
Such approach could include stochastic acceleration (e.g. \cite{1995ApJ...446..699P}),
beam-plasma instability and Langmuir wave generation (e.g. \cite{2013A&A...550A..51H}), electric current associated with the magnetic field twist or
return current (e.g. \cite{2014A&A...561A..72G}),
warm-target effects (\cite{2015ApJ...809...35K}),
or electron anisotropy \cite{2020A&A...642A..79J}.
In Figure \ref{flare-workflow}, step (5) shows the loop model temperature vs. distance along the loop, but at each location (6) the Fokker-Planck code calculates an electron energy distribution; the one shown in this case, $f(E)$, is a thermal plus power-law distribution calculated at the 10 Mm distance.  At a given time $t$, one may place the 1D model as a function of distance $s$ along the loop back into GX Simulator, filling the 3D flux tube by a suitable lateral spread function (Gaussian or other) perpendicular to the central field line.  This populates each of the relevant voxels in the 3D volume with $f(E,\mu)$, the array-defined quantity that can now be used to calculate the microwave emission and absorption along the LOS, and from that perform the  radiative transfer (\S \ref{Sec_Rad_Trans}) to obtain the emergent brightness temperature at each frequency (step 7). Finally, the simulated maps at the resolution of the model must be convolved with the instrumental frequency-dependent point spread function (step 8).  By calculating the convolved emission for many time steps, a multifrequency simulation movie will result that can be compared directly with observations. Currently, we are implementing a transport code module that includes most of the relevant physical processes,
which will compute and return numerical solutions for the flux tubes exported from \gx.

\label{flares}
\section{Radiation Transfer Codes}
\label{Sec_Rad_Trans}
The \gx package provides a series of radiation transfer codes to compute microwave, X-ray, and EUV emission from the same 3D model, which are written directly in IDL, or use an IDL wrapper that calls platform specific external dynamic link libraries (DLL), which are precompiled for WIN32 or WIN64 OS systems,  or shared library codes, which are automatically compiled when first called on Unix, Linux, or MAC platforms.

\subsection{Fast GS and GR Codes\label{fastcodes}}
Initial versions of GX Simulator \citep{Nita2015,Nita2018} included several radio radiation transfer codes obtained from either FORTRAN or C++ source codes developed by \citet{Fleishman2010a,Fleishman2014a}. For the flare case, the corresponding codes included gyrosynchrotron (GS) and free-free emission mechanisms, while for a nonflaring case--gyroresonance (GR) and free-free mechanisms. All codes took into account the mode coupling in the quasitransverse layers and the effect of limiting polarization (only circular, not linear, polarization survives while propagating through the coronal plasma).

In the most recent release, these radio radiation codes have been much improved and generalized. For the nonflaring case, the code performs radiation transfer in a multicomponent multithermal coronal plasma \citep{2021ApJ...914...52F}. This code takes into account chemical composition of the plasma, temperature-dependent ionization states of various elements (He--Zn), exact Gaunt factors, free-free emission due to collisions of thermal electrons with neutral hydrogen and helium, and the effect of magnetic field. In addition, it permits the plasma to be described by DEM/DDM, rather than a single pair of the temperature and density values; this is applied to both the free-free and GR emission calculation. The functionality available in older versions, such as mode coupling and limiting polarization, is preserved in this new release. Currently, this is the most complete and accurate version of the radio radiation transfer code that includes both GR and free-free emission mechanisms.

For the flaring case, \citet{UltimateFastCodes2021} generalized the fast GS codes by \citet{Fleishman2010a} such as the distribution function of the nonthermal electrons can now be defined not only in a simplified analytical form, but also in the form of numerically-defined arrays. Thus, the output of numerical solutions of acceleration/transport models can now be employed by the tool directly. As in the previous versions, the continuous GS approximation provides very high computation speed, while, if necessary, the harmonic structure at low frequencies can be reproduced at some cost of execution time using the exact GS codes. In addition, the free-free component is now treated based on the new theory developed by \citet{2021ApJ...914...52F}, similar to what has been implemented in the nonflaring codes. The calling sequences and interfaces were updated to ease the use of the codes--both within GX Simulator and in standalone applications.

\subsection{X-Ray Codes}
In the current \gx release, the X-ray calculating block (routines \textit{xray\_tt.pro} and \textit{xray\_tt\_albedo.pro})
has been substantially enhanced to include:
i) Hard X-ray calculations using thick-target model \citep{1971SoPh...18..489B};
and ii) X-ray albedo correction (photosphere Compton back-scattered X-rays, see \citet{2006A&A...446.1157K} for details).
In addition, the thermal emission is now calculated using the CHIANTI database\footnote{https://www.chiantidatabase.org}
for plasma temperatures above $0.09$~keV. The default values
use the OSPEX software \citep{2002SoPh..210..165S} default abundances\footnote{https://hesperia.gsfc.nasa.gov/ssw/packages/spex/doc/OSPEX\_explanation.htm}
so that comparison between OSPEX fits of RHESSI data and \gx are readily available.
The threshold temperature for CHIANTI calculations
is controlled by the parameter \textit{Te\_thr=0.09}. The X-ray flux for lower than \textit{Te\_thr} temperatures is
calculated using a simplified, computationally faster, free-free bremsstrahlung expression.
Although the CHIANTI-based method is slower,
this approach provides more precise calculation of soft X-ray emission accounting for free-free, free-bound and line emissions
(see example spectrum in Figure 4.2 \citep{2011SSRv..159..301K}).
Similarly, the soft X-ray calculations are performed using the same routines as in OSPEX, allowing for direct comparisons with RHESSI observations
and to change the default element abundances.

The hard X-ray flux from an emitting volume at a distance $R$ ($\approx1$\,AU in the case of the Sun) is calculated using the mean electron flux \citep{2003ApJ...595L.115B}
\begin{equation}\label{eq:HXR_flux}
I(\epsilon)=\frac{1}{4 \pi R^{2}} \int_{\epsilon}^{\infty} \bar{nVF}(E) Q(\epsilon, E) d E,
\end{equation}
where $\bar{nVF}(E)$ is a mean electron flux integrated along line-of-light voxels. The cross-section $Q(\epsilon, E)$ is approximated by the cross-section used in OSPEX and
tabulated as a look-up table for speed.

To account for both coronal and chromospheric hard X-ray emission, the hard X-ray flux is calculated differently in the chromosphere and the solar corona. The coronal X-ray emission is calculated directly using thin-target emission with $\bar{nVF}(E)$ defined by the modeller, while the chromospheric X-ray emission adopts the thick-target approach.  The electrons penetrating into the chromosphere are used to calculate thick-target emission, so the mean electron flux in  equation \ref{eq:HXR_flux} is calculated as
\begin{equation}
\bar{n V F} (E)=\frac{A}{K} E \int_{E}^{\infty} {F}_{0}\left(E_{0}\right) d E_{0}\,,
\end{equation}
where $K=2 \pi e^{4} \Lambda=2.6 \times 10^{-18} \,\mathrm{cm}^{2}\, \mathrm{keV}^{2}$
and $A$ is the cross-sectional area of the electron beam.

The hard X-ray albedo contribution (angle dependent Green's function correction for photospheric albedo) is computed based on the approach by \citet{2006A&A...446.1157K} and uses pre-calculated Green's matrices used in
OSPEX\footnote{\url{https://hesperia.gsfc.nasa.gov/ssw/packages/spex/doc/ospex_explanation.htm##Albedo\%20Correction}}. Using an anisotropic X-ray electron distribution, hard X-ray fluxes in upward and downward directions are calculated. These fluxes are used to calculate the anisotropic X-ray flux to the observer \cite{2006ApJ...653L.149K,2011A&A...536A..93J,2013SoPh..284..405D}.

\begin{figure}[!htb]
\includegraphics[width=0.45\columnwidth]{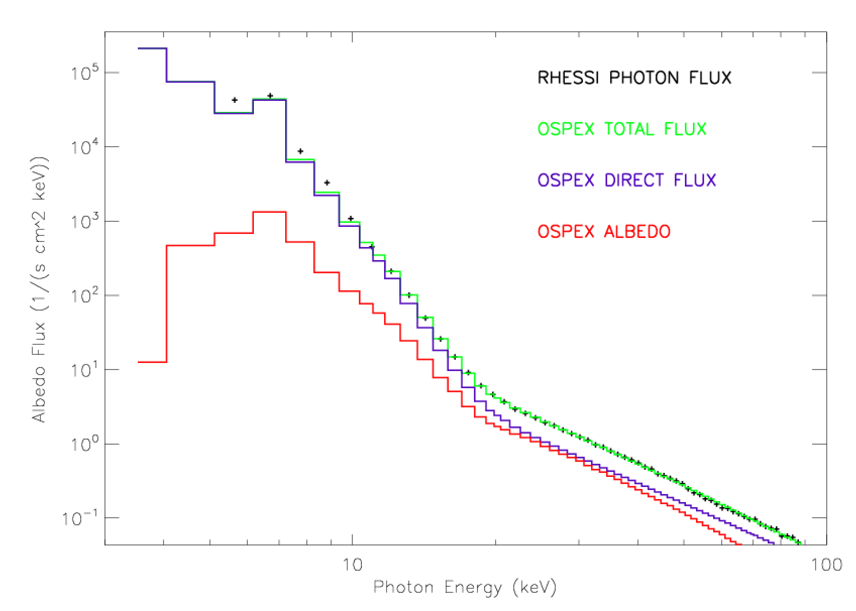}
\includegraphics[width=0.45\columnwidth]{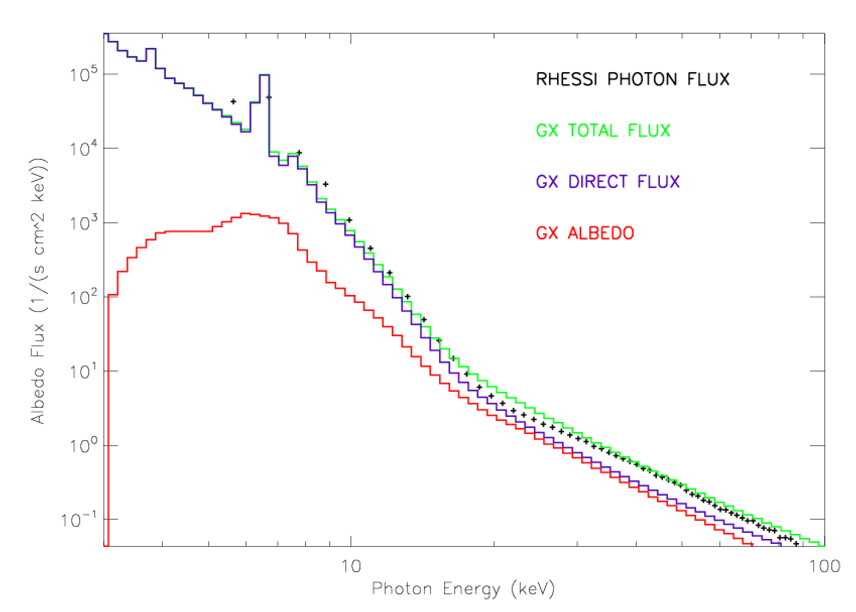}
\caption{\label{fig:albedo} Left panel: OSPEX fit Xray model spectrum for the same instance of the 2017 Sep04 flare used for illustration purposes in Figure \ref{flare-workflow}: observed photon spectrum (symbols), total OSPEX fit spectrum (green), direct OSPEX fit spectrum (blue), OSPEX Albedo contribution (red). The parameters estimated from this fit are:  thermal electrons emission measure: $EM=3\times10^{48}~{cm}^{-3}$; plasma temperature: $T=1.51~{keV}= 1.75\times10^7~{K}$; nonthermal electrons density: $n_{th}=2.3\times10^{34}~{electrons}/{s}$; nonthermal electrons energy index $\delta=3.64$; nonthermal electron minimum cut-off energy : $E_{min}=19.0~{keV}$. Right panel: The model Xray spectrum generated by \gx from a flaring loop model corresponding to the OSPEX fit parameters.}
\end{figure}

For illustration purposes, we show in the left panel of Figure \ref{fig:albedo} the result of the OSPEX fit model spectrum for the same SOL2017-09-04T20:28:00 M5.5 event shown in Figure \ref{flare-workflow}. The parameters estimated from this fit, i.e.  thermal electron emission measure: $EM=3\times10^{48}~{cm}^{-3}$; plasma temperature: $T=1.51~{keV}= 1.75\times10^7~{K}$; nonthermal electron flux: $n_{th}=2.3\times10^{34}~{electrons}/{s}$; nonthermal electron energy index $\delta=3.64$; nonthermal electron minimum cut-off energy: $E_{\min}=19.0~{keV}$ have been used as input parameters for a \gx flaring loop model resulting in the model spectrum shown in the right panel of Figure \ref{fig:albedo}.

\subsection{EUV Codes}
The main routine used by \gx to compute EUV emission from a generic GX model produced by AMPP (\S\ref{section:corona}), is \textit{AIA.pro}, which computes the emission by convolving the DEM distribution from the model with the appropriate temperature response function, $G(T)$, of the SDO/AIA instrument:
\begin{equation}
   I = \int\xi(T)G(T)dT. 
\end{equation}

The functionality of this routine was described by \citet{Nita2018}, who discussed its performance and limitations based on a model to data comparison performed using a model of the AR 11072 on 2010 May 23 and observational SDO/AIA data. A recent upgrade of this rendering routine implemented the use of a time dependent AIA response function that automatically adapts to the time of the model for which the synthetic EUV maps are computed. Currently, \gx does not compute EUV spectra for comparison with EIS or IRIS data; adding this functionality will require an update of the tool.

\subsection{Integration of User-Supplied Radiation Transfer Codes}
The \gx plugin architecture allows straightforward integration of any user-supplied radiation transfer codes, provided that they are interfaced by IDL wrappers that abide by the \gx calling convention given by the following prototype:
\begin{verbatim}
pro generic_wrapper, parms, rowdata, info=info [, $
                    ;optional input/output variables
                    nparms, rparms, user_arg1,user_arg1,..$
                    ;optional input/output keyword variables
                    user_keyword1=user_keyword1,..]
                    \end{verbatim}
As indicated above, the user defined IDL wrapper must accept three mandatory arguments: the strictly ordered and strictly named \bfemph{parms} and \bfemph{rowdata} variables, and the \bfemph{info} keyword. In addition, the wrapper may accept two optional, strictly named input variables, \bfemph{nparms} and \bfemph{rparms}, as well as an arbitrary number of optional user defined variables and/or keywords that may be used to pass internal user defined data between subsequent calls of the wrapper that happen during the same geometrical scan of the model.

The use of the mandatory \bfemph{parms} input variable is reserved for passing to the wrapper the model parameters corresponding to a two dimensional geometrical slice of the 3D volume along the LOS direction, in a form of a $N_x\times N_z \times N_{parms}$ double precision numerical array, where $N_x$ is the number of image pixels along one row of the synthetic map image, $N_z$ is the number of nodes along any given line of sight, and $N_{parms}$ the number of model properties needed to solve the radiation transfer equation.

Starting with the current release of the \gx package, the optional, strictly named \bfemph{nparms} and \bfemph{rparms} have been introduced to allow more memory efficient passing of integer or, respectively, floating point model or setup parameters that are LOS-independent.

The use of the mandatory \bfemph{rowdata} output variable is reserved for retrieving from the wrapper the multidimensional output data corresponding to one row of synthetic image, in a form of multidimensional floating point array that is expected to have at least two reserved dimensions, $N_x\times N_{chan}$, where $N_{chan}$ represents the number of data channels to be computed, such as, for example, the number of microwave frequencies, the number of X-ray energy channels, or the number of EUV/UV passbands. However, the \bfemph{rowdata} output variable may have additional user defined dimensions, which may be used to accommodate, for example, different Stokes parameters and/or other radiation-mechanism-specific data channels.

The \bfemph{info} mandatory keyword  must be used by the wrapper to return, on request,  metadata information describing the expected model parameters and the structure of the radiation transfer data output, in a form of an IDL structure. This structure contains a set of mandatory tags that are needed by \gx to:
\begin{itemize}
\item retrieve the information needed to initialize the input parameter arrays described above and dynamically link them to the expected model data
\item dynamically generate the wrapper specific user interface input fields needed to customize the options of the selected wrapper
\item customize the memory space and data visualization displays needed to hold and inspect the image cubes generated by the selected radiation transfer code.
\end{itemize}

Since a full description of the underlying \gx architecture related to the radiation code calling conventions is beyond the scope of this paper, we refer the interested reader to inspect the IDL wrappers already included in the \gx package, which may be used as templates for designing customized wrappers.

\section{Conclusions}
The \gx\ tool is now mature enough to address various modeling needs in both flare and AR science. It offers convenient automatic ways to build 3D models, fine tune them, generate synthetic observables, perform data-to-model comparison, and, eventually produce 3D models compatible with all available observational constraints.

\begin{acknowledgments}
The study is supported by NSF grants
AGS-2121632,  
AST-2206424,  
AGS-1743321, 
AGS-2130832, 
and NASA grants
80NSSC20K0718 
80NSSC20K0627, 
80NSSC19K0068, 
and 80NSSC23K0090, 
to New Jersey Institute of Technology. S.J.S.'s contribution was supported by an appointment to the NASA Postdoctoral Program at the Goddard Space Flight Center, administered by Universities Space Research Association under contract with NASA.
\end{acknowledgments}

\appendix
\section{GX SIMULATOR Automatic Model Production Pipeline IDL Script Example\label{appendix:a}}
Here we list an AMPP script that may be used to generate a series of POT, BND, NAS, GEN and CHR models (as described in Table \ref{tab:file_types}) for the AR11520.
\begin{verbatim}
\input{ampp.bat}
\end{verbatim}

At run-time, the script listed above  prints out a set of console messages that may be used to to assess the system performance. Here we list the console messages generated by this AMPP script when run on a Windows 10 system equipped with an  Intel Xeon E-2286M CPU 2.4 GHz, 8 cores, 64 GB RAM.

\begin{verbatim}
;IDL console messages
% GX_FOV2BOX: Potential extrapolation performed in 4.499 seconds
% GX_FOV2BOX: Bound Box structure saved to
C:\gx_models\2012-07-12\hmi.M_720s.20120712_044626.W82S16CR.CEA.BND.sav
% GX_FOV2BOX: Performing NLFFF extrapolation
% GX_FOV2BOX: NLFFF extrapolation performed in 249.658 seconds
% GX_FOV2BOX: NLFFF box structure saved to
C:\gx_models\2012-07-12\hmi.M_720s.20120712_044626.W82S16CR.CEA.NAS.sav
% GX_FOV2BOX:
Computing field lines for each voxel in the model..
% GX_ADDLINES2BOX: Field line computation performed using DLL
implementation in 104.818 seconds
% GX_FOV2BOX: Box structure saved to
C:\gx_models\2012-07-12\hmi.M_720s.20120712_044626.W82S16CR.CEA.NAS.GEN.sav
% GX_FOV2BOX: Generating chromo model..
% GX_FOV2BOX: Chromo model generated in 9.248 seconds
% GX_FOV2BOX: Box structure saved to
C:\gx_models\2012-07-12\hmi.M_720s.20120712_044626.W82S16CR.CEA.NAS.CHR.CHR.sav
% GX_FOV2BOX: This script generated the following files:
% GX_FOV2BOX:
C:\gx_models\2012-07-12\hmi.M_720s.20120712_044626.W82S16CR.CEA.BND.sav
C:\gx_models\2012-07-12\hmi.M_720s.20120712_044626.W82S16CR.CEA.NAS.sav
C:\gx_models\2012-07-12\hmi.M_720s.20120712_044626.W82S16CR.CEA.NAS.GEN.sav
C:\gx_models\2012-07-12\hmi.M_720s.20120712_044626.W82S16CR.CEA.NAS.CHR.sav
\end{verbatim}

\section{IDL script to generate a customized EBTEL coronal model of an active region and associated synthetic microwave maps \label{appendix:b}}
Here we present a script that may be used to customize the EBTEL coronal model of the \emph{hmi.M\_720s.20120712\_044626.W82S16CR.CEA.NAS.CHR.sav} AR11520 model produced by the AMPP script presented in the previous section and generate, for a selected field of view, synthetic microwave maps at two frequencies, 5.7GHz and 17GHz, matching the onserving frequencies of the Siberian Solar Radio Telescope (SSRT) and Nobeyama Radio Heliograph (NoRH) radio interferometry arrays.
\begin{verbatim}
input{chmp.bat}
\end{verbatim}
The run-time console messages generated by this AMPP script when run on a Windows 10 system equipped with an  Intel Xeon E-2286M CPU 2.4 GHz, 64 GB RAM are listed below.
\begin{verbatim}
;IDL console messages
% GX_TBMAPS_SCRIPT: Map object saved to c:\moddir\maps\test.map
% GX_TBMAPS_SCRIPT: GXCUBE data structure saved to c:\moddir\gxc\test.gxc
% GX_TBMAPS_SCRIPT: Synthetic maps computed in 125.231 seconds
\end{verbatim}

\section{Simple steps to generate  GX SIMULATOR--compatible Density-Temperature models\label{appendix:c}}
To populate the volume of the \gx compatible magnetic field model with numerical thermal plasma density and temperature pairs produced by alternative means, one may start from a standard \textbf{.GEN.} IDL structure, which contains three specific one-dimensional, same-size array tags, namely \textbf{IDX}, \textbf{LENGTH}, and \textbf{BMED}, where \textbf{IDX} is used to store the one-dimensional indices of the coronal voxels crossed by closed magnetic field line having the lengths and averaged magnetic fields stored in the other two arrays.
One possible, straightforward approach would be to add to such standard box structure two additional, same-size array tags, namely \textbf{N} and {T} that would store the $n$,$T$ pairs defined in exactly the same volume voxels referenced by the \textbf{IDX} tag. If this approach is chosen, the \gx user would be offered the choice to synthesize multi-wavelength thermal coronal emission directly from such $n$,$T$ distributions, while still preserving the choice to use, as an alternative, the EBTEL approach described in \S\ref{section:corona} to compute and use the DEM-derived $n$,$T$ pairs provided by Equation \ref{n-T}.

However, the \textbf{.GEN.} structure lists only voxels that are crossed by closed magnetic field lines within the model box. To assign $n,T$ pairs to an unlimited set of voxels in the box, one may instead start from a standard \textbf{.POT.} or \textbf{.NAS.}, or compatible, structure to which the same-size array tags \textbf{IDX} and \textbf{N} and \textbf{T} may be added, in which case \textbf{IDX} may reference the entire volume, or any portion of it.


\end{document}